\documentclass[twocol]{ametsocV5}

\usepackage{amsmath,amsfonts,amssymb,bm}
\usepackage{mathptmx}%{times}
\usepackage{newtxtext}
\usepackage{newtxmath}

\title{On the Asymmetry of Rotating Stratified Throughflows Across Finite Amplitude Topography. Part I: Injection of Boundary Layer PV}

\author{Miguel A. Jim\'{e}nez-Urias\correspondingauthor{Miguel Jim\'{e}nez-Urias, mjimen17@jh.edu}}

\affiliation{Earth and Planetary Sciences, Johns Hopkins University, Baltimore, MD}

\extraauthor{LuAnne Thompson}
\extraaffil{School of Oceanography, University of Washington, Seattle, WA}

\abstract{We present an idealized study of the baroclinic structure of a rotating, stratified throughflow across a finite amplitude ridge in the f-plane that is forced by a steady, uniform inflow and outflow of equal magnitude. The resulting equilibrated circulation is characterized by unstable, western-intensified boundary currents and flow along f/H contours atop the ridge near the crest that closes the forced circulation. We find that bottom Ekman dynamics localized to lateral boundary currents amplify the stratification expected by simple stretching/squashing: a strongly stratified bottom front (high PV anomaly) along the anticyclonic boundary current associated with net upslope transport, and a bottom mixed layer front (vanishing PV anomaly) localized to the cyclonic boundary current where there is net downslope transport. PV anomalies associated with the two fronts are advected by both the mean flow and eddies that result from baroclinic growth, resulting in a spatial distribution where high PV is concentrated along the ridge, and low PV is advected into the interior ocean downstream from the ridge, at mid-depth. Using a framework of volume integrated PV conservation which incorporates the net fluxes associated with bottom topography, we conform an approximate integral balance between the PV injection from the bottom boundary layer on the ridge and net advection across the ridge. Implications of these findings for understanding the interplay between large scale and bottom boundary dynamics are discussed.}

\begin{document}

\maketitle

\section{Introduction}

The vertically integrated pathway of topographically steered, rotating flows across finite amplitude topography can be understood from integral constraints on the potential vorticity (PV) of the fluid \citep{yang2000water, yang2007potential}, where topographic contours represent free streamlines and PV dissipation associated with boundary currents is balanced by net advective PV fluxes that arise from cross-slope transport. Thus, the expected circulation of a throughflow across a ridge is that of an anticyclonic western boundary current as the flow circulates from the deep ocean into the ridge crest, and a cyclonic, eastern boundary current (pseudo westerward intensification) as the flow circulates into deeper water downstream from the ridge. Along the ridge crest, the flow is expected to follow $f/H$ contours thus closing the circulation by connecting the two boundary current systems. Exchanges of watermass properties between ocean basins can be model as topographically steered, forced flows that are constrained laterally by abrupt coastal shelfs, such as the flow of North Atlantic waters across the Iceland-Faroe Ridge \citep{hansen2003iceland, osterhus2005measured,hansen2008inflow,hansen2010stability}, and even exchanges between marginal seas like the throughflow in the South China Sea \citep{qu2006south,qu2009introduction}.

An overlooked component of boundary currents (and therefore throughflows) is their vertical baroclinic structure, in particular the secondary ageostrophic circulation associated with the bottom boundary layer \citep{pedlosky1968overlooked}. Depending on the orientation of the mean (interior) flow with respect to f/H contours, bottom friction can lead to a cross-slope Ekman transport that rapidly modifies the stratification within the boundary layer, and that of the interior flow if it results in a convectively unstable stratification \citep{condie1995descent, waahlin2001downward,brink2010buoyancy}. If the along-slope interior flow experiences little variability and the transit time along the ridge is sufficiently long, bottom friction can lead to an Ekman arrest, when the Ekman transport promotes a horizontal buoyancy gradient in thermal wind balance such that the geostrophic shear reduces the effects of the bottom boundary onto the mean interior flow via \textit{buoyancy shutdown} \citep{maccready1991buoyant, trowbridge1991asymmetric,maccready1993slippery, ruan2016bottom}. Furthermore, baroclinicity within a \textit{thick} bottom boundary layer can lead to baroclinic instability, thus a rapid restratification of the bottom boundary layer \citep{wenegrat2018submesoscale}. Lastly, diapycnal mixing at the bottom can drive frictional flow up the slope by weakening the stratification, thus promoting a stable boundary layer \citep{wunsch1970oceanic, phillips1970flows,thorpe1987current, benthuysen2012friction}. 

In the case of throughflows across finite amplitude ridge topography with channel geometry, the flow can experience significant along stream variability at it transits the ridge. Since the transit time is not infinite, it is not apparent if the along-slope flow will experience Ekman arrest (buoyancy shutdown). Nonetheless, Ekman dynamics within a thick boundary layer can modify the upper interior ocean through rapid re- or de-stratification resulting in high or low bottom PV anomalies, respectively. Then, depending on the resulting localized distribution of the PV within the bottom boundary layer, the mean flow or baroclinic eddies can advect such PV anomalies away into the interior. Such non-local redistribution by eddies has already been explored in idealized studies of subduction in surface fronts \citep{spall1995frontogenesis, manucharyan2013generation} and shelf break boundary currents \citep{spall2008western}, although in these studies, bottom boundary layer dynamics are largely unresolved and the flows are not equilibrated.

%Similarly, most studies of bottom boundary layer dynamics do not consider equilibrated interior flows with significant along-stream variability, and thus assume that the two-dimensional plane view dynamics equally apply along the infinitely long, along-slope direction. Instead, along slope variability is likely to induce a geography of diapycnal mixing, and potentially a non-local modification due to advective fluxes.

How bottom boundary layer control may come about in topographically steered throughflows is unclear, particularly when the mean flow experiences significant along-stream changes as it navigates the ridge. That is, along-stream variability can lead to localized patterns of bottom diapycnal mixing, while advective fluxes can redistribute PV leading to non-local effects.

In this study we investigate the effects associated with baroclinic eddies and bottom boundary layer dynamics on the throughflow across a finite amplitude ridge using a combination of theory and numerical simulations of the primitive equations. In particular, we analyze the Ertel PV of the fluid, a dynamically active tracer with conservation properties that incorporate both thermodynamic and dynamical aspects of the flow \citep{haynes1987evolution, haynes1990conservation}. What makes ideal the use of the PV of the fluid, is its ability to describe density and mean flow anomalies as \textit{High} or \textit{Low} PV with respect to background values, useful when looking at the PV along the path of the flow (streamlines) or along isentropes (surfaces of constant temperature). In an accompanying paper, we study the effect these processes have on the vertically integrated circulation.

The outline of the paper is as follows: In section \ref{secPV_theory} we derive expressions for boundary PV fluxes that generalize the integral PV balance, including a term associated with the presence of bottom topography. We split the topographic PV terms in the integral balance into those evaluated at lateral walls, and those along f/H contours. In section \ref{model_conf} we outline the model configuration for all simulations, and then in section \ref{results} we present the main results beginning with a description of the vertical structure of the mean fields associated with equilibrated simulations, maps of PV along isentropes and flow fields along streamlines, in order to interpret changes the throughflow experiences as it navigates the ridge. We then present an integral PV balance, i.e. boundary PV sources to complement the previous analysis. Lastly, we summarize our findings and conclusions in section \ref{d_and_c}.

\section{Boundary sources of potential vorticity}\label{secPV_theory}

In the primitive equations, the conservation of (Ertel) potential vorticity $q=\bm{\omega}_{a}\cdot\nabla b$ in flux form is given by

\begin{equation}\label{PV_eqn}
\frac{\partial q}{\partial t} = -\nabla\cdot\mathbf{J}
\end{equation}

where $\bm{\omega}_{a}=(-v_{z},u_{z}, f+\zeta)$ is the absolute vorticity with $\mathbf{u}=(u,v,w)$ the 3D velocity, $\zeta=v_{x}-u_y$ the relative vorticity and $b=g\alpha T$ the buoyancy of the fluid (where we have assumed a linear equation of state, $T$ being the temperature). The PV flux vector can be written in two equivalent forms

\begin{equation}\label{Jvector}
\mathbf{J}=\underbrace{\mathbf{u}q}_{\mathbf{J_{A}}}+
\underbrace{\nabla b\times\mathbf{F}}_{\mathbf{J_{NA}}}= \nabla b\times\nabla B + \nabla b \times\left(\frac{\partial\mathbf{u}_{h}}{\partial t} \right) - \bm{\omega}_{a}\frac{\partial b}{\partial t}
\end{equation}

$B=(1/2)|\mathbf{u}|^2+g(\eta-z)+p/\rho_{0}$ is the Bernoulli potential, where  $\eta$ is the surface elevation, $p$ the fluid pressure due to stratification, and $\mathbf{F}=\mathbf{F}_{h}+\rho^{-1}_{0}\partial_{z}\bm{\tau}$ represents dissipation of horizontal momentum, with $\mathbf{F}_{h}$ being the viscous, harmonic horizontal dissipation of momentum.

The two equivalent formulations of the PV flux vector $\mathbf{J}$ in (\ref{Jvector}) allow complementary interpretations: $\mathbf{J}$ can be decomposed into an advective and a dissipative flux contribution \citep{haynes1987evolution, haynes1990conservation}, and so in the absence of dissipative or diabatic processes, the evolution of (Ertel) potential vorticity is given entirely by the divergence of advective fluxes. Furthermore, the advective PV flux can, in the presence of eddying variability, be split into mean and eddy fluxes, $\textit{i.e.}$ $\mathbf{J}_{A}=\overline{\mathbf{u}}\overline{q} + \overline{\mathbf{u}'q'}$, where the overbar $\overline{()}$ represents time-mean, and primed variables $()'$ represent the deviation (eddy) from such mean. Lastly, if the flow is steady (last two terms in (\ref{Jvector}) are zero), the intersections of surfaces of constant Bernoulli potential and surfaces of constant buoyancy/temperature represent streamlines for the total potential vorticity flux \citep{schar1993generalization, bretherton1993flux}. 

In the quasigeostrophic limit, the intersection of temperature/buoyancy surfaces with bottom topography can be represented as PV-delta sheets \citep{bretherton1966critical,rhines1979geostrophic}, which can be tapped by the circulation and be advected by the mean flow, leading to downstream formation of cyclonic anomalies \citep{hallberg2000boundary,schneider2003boundary}. In what follows, we derive a local flux formulation of the topographic PV flux in primitive equations that is general enough that it can incorporate dynamics beyond QG, and allow us to diagnose the topographic PV flux that is consistent with a volume integrated \textit{Eulerian} Ertel PV balance.

Using the Bernoulli formulation of the PV flux vector in (\ref{Jvector}), the topographic PV flux can be written as the sum of two contributions as follows

\begin{equation}\label{two_decomp}
\overline{\mathbf{J}}\cdot\mathbf{n}_{bot} = \overline{\mathbf{J}}^t\cdot\mathbf{n}_{bot}+\overline{\mathbf{J}}^s\cdot\mathbf{n}_{bot}
\end{equation}
 
where the superscripts $t$ and $s$ differentiate between the fluxes that explicitly depend on the \textit{tendency} of flow variables (the last two terms in \ref{Jvector}), and those that do not. $\mathbf{n}_{bot}=(0,-dh/dy,-1)$ is the local normal vector to bottom topography, outward pointing. The terms in (\ref{two_decomp}) are

\begin{equation}\label{Jtbot}
\overline{J}^t_{bot}=\mathbf{n}_{bot}\cdot\left[\overline{\nabla b\times\left(\frac{\partial\mathbf{u}_{h}}{\partial t}\right)} - \overline{\bm{\omega_{a}}\frac{\partial b}{\partial t}}\right]\bigg\rvert_{z=-h}
\end{equation}

and

\begin{equation}\label{Jsbot}
\overline{J}^s_{bot}=\mathbf{n}_{bot}\cdot\left(\overline{\nabla b\times\nabla B}\right)\big\rvert_{z=-h}
\end{equation}

In the presence of finite amplitude topography, linearization of the bottom boundary condition is no longer possible. Thus we introduce a $\sigma$-coordinate transformation to represent a horizontal (\textit{i.e.} meridional) derivative along terrain-following coordinates \citep{shchepetkin2005regional}. The terrain-following derivative is

\begin{equation}\label{sigma_transform}
\frac{\partial (\cdot)}{\partial y}\bigg\rvert_{\sigma=-1} = \bigg(1-\gamma_{(\cdot)}\bigg)\frac{\partial (\cdot)}{\partial y}\bigg\rvert_{z=-h}
\end{equation}

where $\sigma=\sigma(\eta,h)$ is the terrain following coordinate with a nonlinear \textit{stretching} function, such that $\sigma(0)=\eta(x,y, t)$ and $\sigma(-1)=-h(y)$.
The left hand side of (\ref{sigma_transform}) represents an along-bottom derivative of a scalar function denoted by $(\cdot)$, where $\gamma_{(\cdot)}=(dh/dy)\big/((\cdot)_{y}/(\cdot)_{z})$ is the ratio of local slope of scalar $(\cdot)$ to the topographic slope, which vanishes for flat bottom topography. $\gamma_{(\cdot)}$ arises as a \textit{stability} \textit{slope} parameter that measures the effect of the sloping bottom on baroclinic instability, when considering the scalar buoyancy ($\gamma_{b}$) \citep{blumsack1972mars, lozier2005influence, isachsen2011baroclinic}. 

Given the transformation (\ref{sigma_transform}), the (time-mean) contribution of (\ref{Jtbot}) is

\begin{equation}\label{Jtbot3}
\overline{J^{t}_{bot}}=\underbrace{-\mathbf{\hat{k}}\cdot\left(\overline{\nabla_{\sigma}b\times\frac{\partial\mathbf{u}_{h}}{\partial t}}\right)\bigg|_{\sigma=-1}}_{\overline{J}^{t}_{b1}}+\underbrace{\bigg[\overline{\big(f+\zeta_{\sigma}\big)\frac{\partial b}{\partial t}}\bigg]\bigg|_{\sigma=-1}}_{\overline{J}^{t}_{b2}}
\end{equation}

where $\mathbf{\hat{k}}$ is the unit vector perpendicular to depth-surfaces. Thus $\overline{J}^{t}_{bot}=0$ either when the flow variables are steady, or when surfaces of constant buoyancy (temperature) do not intersect the bottom ($\nabla_{\sigma}b=0$ and so $\overline{J}^{t}_{b1}=0$), and when the along-bottom vorticity $\zeta_{\sigma}=-f$ (so $\overline{J}^t_{b2}=0$).

The contribution to the net topographic PV flux by the second term term in (\ref{two_decomp}) can be written as

\begin{equation}\label{sJbot2}
\overline{J}^{s}{_{bot}}=\bigg[\overline{B_{x}b_{y}\left(1- \gamma_{b} \right)} - \overline{b_{x}B_{y}\left(1-\gamma_{B}\right)}\bigg]_{z=-h(y)}
\end{equation}

or, using $\sigma$-coordinates,

\begin{equation}\label{sigmaPVbot}
\overline{J}^s_{bot}=\overline{\dfrac{\partial_{\sigma}(B,b)}{\partial(x,y)}}\bigg\rvert_{\sigma=-1}
\end{equation}

where the subscripts $\sigma$ on the Jacobian term $\partial_{\sigma}(\cdot,*)/\partial(x,y)$ implies that derivatives are along sigma (terrain following) coordinates.

Now, we introduce the functional $\Gamma_{b}=\Gamma[b;y,z]$ acting on the scalar function $b$ (although it can be any scalar function), defined as

\begin{equation}
\Gamma_{b}=\left(\dfrac{\partial b}{\partial y}\bigg|_{\sigma=-1}\right)\bigg/ \left(\dfrac{dh}{dy}\right)
\end{equation}

The functional $\Gamma_{b}$ represents the ratio between along-bottom northward gradient (\textit{i.e.} along $\sigma=-1$ surface) to topographic slope, where we have used the transformation (\ref{sigma_transform}) for simplicity and compact notation. With this, the expression in (\ref{sJbot2}) is written compactly as

\begin{equation}\label{sJbot2n}
\overline{J^{s}}_{bot}=\bigg[\overline{B_{x}\Gamma_{b}} -\overline{b_{x}\Gamma_{_{B}}}\bigg]\bigg\rvert_{z=-h(y)}\frac{dh}{dy}
\end{equation}

We now decompose $\overline{J}^s_{bot}$ into contributions from (lateral) boundaries (\textit{i.e.} evaluated where topography intersects lateral boundaries) and terms along $f/H$ contours. Then, the time-mean net topographic PV flux $\overline{J}_{bot}$, given by the sum (\ref{two_decomp}) is 

\begin{equation}\label{sJbot4n}
\overline{J}_{bot}= \Delta_{x}\overline{J}^{*}_{bot} + \int\int\left(\overline{b_{bot}\frac{\partial\Gamma_{_{B}}}{\partial x}}-\overline{B_{bot}\frac{\partial\Gamma_{b}}{\partial x}}\right)\bigg\rvert_{z=-h}\frac{dh}{dy}dxdy + \int\int \overline{J}^{t}_{bot}dxdy 
\end{equation}

where $\Delta_{x}(\cdot) = (\cdot)_{x=L} - (\cdot)_{x=0}$ is the cross-channel difference of $(\cdot)$ (\textit{i.e.} difference between the evaluation at eastern and western walls). The first term on the RHS of (\ref{sJbot4n}) is associated with effects on lateral walls (\textit{e.g.} form stresses) and vanishes in a zonally periodic domain. The second term in (\ref{sJbot4n}) is associated with changes of buoyancy, pressure and kinetic energy across and along f/H contours, while the third term is associated with the $\textit{tendency}$ terms in (\ref{Jtbot3}). 

The first term on the RHS of (\ref{sJbot4n}) can be expanded as follows:

\begin{equation}\label{Jbot1}
\Delta_{x}\overline{J}^{*}_{bot} = \Delta_{x}\overline{\mathcal{F}^{BT}_{_{\Gamma}}} +  \Delta_{x}\overline{\mathcal{F}^{BC}_{_{\Gamma}}}+ \Delta_{x}\overline{\mathcal{E}_{_{\Gamma}}}-\Delta_{x}\overline{\mathcal{S}_{_{{\Gamma}}}}
\end{equation}

where

\begin{eqnarray}\label{Lateral_PV_terms}
\mathcal{F}^{BT}_{_{\Gamma}} & = &\int_{0}^{M}(g\eta\Gamma_{b})\big\rvert_{z=-h}\frac{dh}{dy}dy\nonumber\\ 
\mathcal{F}^{BC}_{_{\Gamma}} & = & \int_{0}^{M}\left(\frac{p\Gamma_{b}}{\rho_{0}}\right)\bigg\rvert_{z=-h}\frac{dh}{dy}dy \nonumber\\
\mathcal{E}_{_{\Gamma}} & = & \frac{1}{2}\int_{0}^{M}(\Gamma_{b}v^2)\big\rvert_{z=-h}\frac{dh}{dy}dy \nonumber\\
\mathcal{S}_{_{\Gamma}} & = & \int_{0}^{M} (b\Gamma_{_{B}})\big\rvert_{z=-h}\frac{dh}{dy}dy\nonumber\\
\end{eqnarray}

The first three terms in (\ref{Lateral_PV_terms}) are associated with the bottom Bernoulli potential, and in particular the first two integral terms represent weighted form stresses (weight being $\Gamma_{b}$) associated with flow separation from the lateral walls. Each integral in (\ref{Lateral_PV_terms}) vanishes when the integrand is symmetric with respect to the (symmetric ridge). However, it is the difference between the wall-contributions (\ref{Jbot1}) that results in net PV fluxes.\\

Our formulation of net topographic PV fluxes (\ref{sJbot4n}) shows the role of form stresses associated with lateral boundary currents in promoting local (and net) PV fluxes (\ref{Jbot1}), as well as the potential role of along-slope variations of buoyancy and bottom Bernoulli potential in promoting local and integral PV fluxes (second integral terms in \ref{sJbot4n}). Along slope variations of buoyancy and bottom Bernoulli potential are associated with viscous dissipation, ageostrophic flows an eddies. The question of which boundary PV terms make the largest contribution, the location of maxima or minima along f/H contours, and what effect these flux have on the mean fields associated with an equilibrated throughflow past finite amplitude topography is examined in subsequent sections.

\section{Model Configuration}\label{model_conf}
\begin{figure}
    \centering
    \includegraphics[width=425pt]{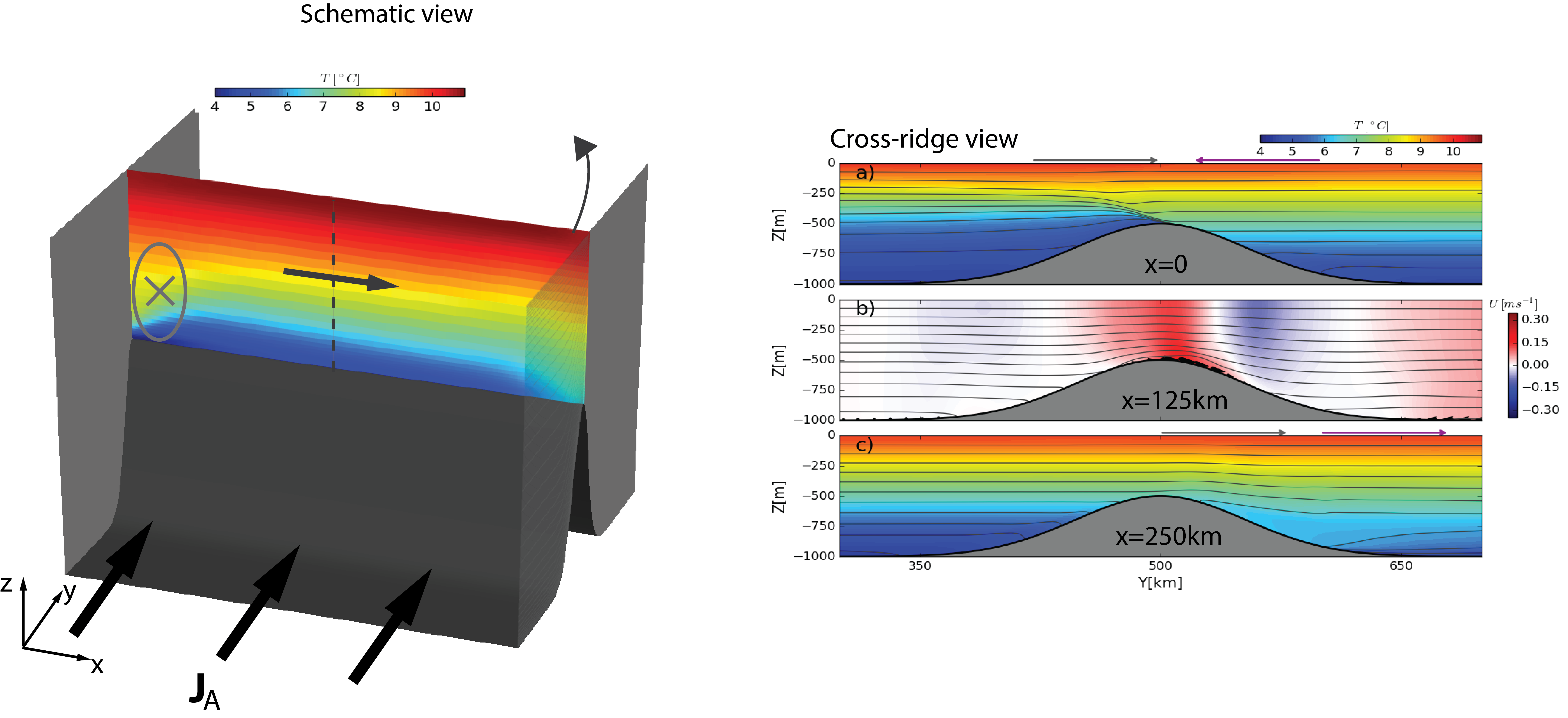}
    \caption{Left: 3D flow configuration showing the vertical stratification of atop the ridge crest in equilibrated simulation $cSt_{ns}$ (we find a similar flow pattern in all stratified simulations). Thick black arrows display the orientation of the inflow at the southern boundary. Gray arrows and circled cross represent the general pathway of the mean flow. a) Mean stratification $\overline{T}$ at the western ($x=0$) wall. b) Mean zonal flow ($\overline{u}$) at the centerline of the channel, shown in the schematic view as a dashed vertical line. Black arrows near the bottom represent the (frictional) bottom flow. c) Mean stratification at the eastern wall ($x=250$km). Gray arrows atop b) and c) display the direction of the flow, with the magenta arrows the direction of propagation of Kelvin waves.}
    \label{fig:diagram}
\end{figure}

We use the Regional Ocean Modeling System (ROMS) that solves the primitive equations using a terrain-following vertical coordinate \citep{shchepetkin2005regional}.
The rectangular channel domain has dimensions $L=250$km and $M=1000$km in the along-ridge ($x$) and across-ridge ($y$) directions. Bottom topography is characterized by a (symmetric) ridge topography $h=h(y)$ that intersects the lateral walls, with a minimum channel depth of $500$m at the crest ($h_{max}=500m$), and maximum channel depth of $H_{0}=1000$m. The resolution in the vertical grid varies as a function of the distance from the ridge, and the top and bottom boundaries. We use 40 vertical levels that allow for a maximum resolution of $dz\approx3\;$m near the sea surface and ocean bottom over the ridge crest, and a minimum resolution of $dz=32\;$m at intermediate depths, away from ridge topography and vertical boundaries. The horizontal resolution is constant, with $\Delta x=1\;$km and $\Delta y=2\;$km. 
The Coriolis parameter is kept constant at $f=1.25\times10^{-4}s^{-1}$, and the model is forced by a lateral uniform volume inflow of $Q_{in}\approx3\;$Sv (1 Sv=$10^6 m^{3}s^{-1}$), an outflow $Q_{out}$ of equal magnitude. A sponge layer is applied within the nudging region to further bring the boundary condition to that of the imposed inflow/outflow. Furthermore, the inflow and outflow are relaxed towards a background stratification $N_{0}^2= 1\times10^{-5}s^{-2}$ every 5 days, the relaxation timescale decaying linearly within 10 grid points of the open boundaries. The combination of volume flux, sponge layer and relaxation make it so that the flow is forced by a (purely) advective PV flux $\mathbf{J}_{_{A}}= N_{0}^2v_{0}f \mathbf{\hat{j}}\approx1.5\times10^{-11}ms^{-4}$ at both the inflow and outflow locations (see Fig \ref{fig:diagram}).

The model dissipates momentum laterally via a harmonic viscous term with an amplitude defined by the coefficient $A_{_{H}}$. The model employs a quadratic bottom drag with drag coefficient $C_{D}=2.5\times10^{-3}$, and we implement the KPP parameterization for (nonlocal) vertical mixing \citep{large1994oceanic}. We vary $A_{_{H}}$ for each different experiment, which has the effect of varying the width of the lateral boundary currents given by the scaling $\lambda_{_{M}}=(A_{h}/\beta_{_{T}})^{1/3}$, where $\beta_{_{T}}=(f/h^2)dh/dy$ is the topographic beta (see Table \ref{table_one}). We measure the non-linearity of the (lateral) boundary currents through the boundary current Reynolds number $Re=V_{_{max}}\lambda_{M}/A_{_{H}}$, a measure of the relative role of inertial and viscous dynamics. While we considered simulations with different values of $N_0^2$, as well as different topographic heights ($h_{max}=350m$ and $h_{max}=650m$), we observed no significant dynamical differences due to the strong barotropic nature of the simulations. The analysis takes place $100$km away from the nudging regions, localized to northern and southern open boundaries. We consider both free-slip and no-slip tangential boundary conditions resulting in the flow configurations which flux momentum laterally (no-slip), or not (free-slip), across lateral walls. We also consider simulations without stratification, as reference simulations, but these are only considered in the accompanying paper, which examines the vertically integrated circulation (See Table \ref{table_one} for parameters associated with mean flow circulation across large amplitude topography). Each simulation is ran for 7 years, of which the first 12 months represent the spin up, a relatively short time due to the strongly barotropic nature of the simulations. 

Given the potential role of the bottom boundary layer on the along-slope flow far away from the influence of lateral boundary currents, we introduce relevant parameters that arise in the two-dimensional, (vertically) semi-infinite bottom boundary layer theory (see Table \ref{table_two}). A parameter of great importance of along slope flows is the (\textit{small angle}) slope Burger number $S\approx (N\theta/f)^2$ where $\theta\approx dh/dy$. The timescale at which buoyancy forces balance an (initial) up/down-slope Coriolis force is given by the buoyancy shutdown timescale, defined as $\mathcal{T}_{shut}=P_{r}^{-1}S^{-2}f^{-1}$, valid for $S\ll1$ (here $Pr\approx1$). Similarly, the stratified frictional spindown time scale is given by $\mathcal{T}_{spin}=E^{-1/2}f^{-1}$, where $E=2A_{\nu}/fH^2_{p}$ is the Ekman number and $H_{p}=fL/2\pi N$ the Prandtl depth, the depth scale over which anomalies penetrate into the interior. The inertial and diffusive time scales are expressed as $\mathcal{T}_{inertial}= f^{-1}$ and $\mathcal{T}_{diffusive}=E^{-1} f^{-1}$. An additional time scale to consider, is the \textit{transit} time scale along $f/H$ contours, defined as an advective time scale $T_{adv}=L_{w}/\overline{U}$, where $L_{w}$ is a lateral scale along the ridge away from the influence of lateral boundary currents, and $\overline{U}$ a scale of the velocity field there. Using $L_{w}\approx 100$km, and a typical velocity $\overline{U}\approx 0.3m/s$, then $T_{adv}\approx4$days (see Table \ref{table_two}).

The diffusive boundary layer has thickness of $\delta_{T}=\left(2A_{\kappa}\mathcal{T}_{diff}\right)^{1/2}$, whereas the bottom boundary layer thickness defined by $\delta_{Ek}=\sqrt{2A_{\nu}/f}$ (where $A_{\nu}$ is not the background value, but rather a model output variable) can vary spatially according to patterns of enhanced vertical (turbulent) mixing and potentially due to eddy variability. Eddies that propagate along f/H contours can introduce temporal variability on the thickness of the bottom boundary layer, measured by the ratio $r_{\omega}=\omega_{bbl}/f$, where $\omega_{bbl}$ is the frequency of the oscillation (eddies) within the bottom boundary layer \citep{ruan2016bottom}.

\section{Results}\label{results}

In this section, we analyze simulations of continuously stratified, rotating throughflows under channel geometry and characterized by a net mean transport across finite amplitude bottom topography. Throughout this paper, unless otherwise stated, we focus primarily on simulation $CSt_{ns}$ characterized by narrow ($\lambda_{_{M}}\approx 17km$) unstable lateral boundary current and $Re\sim O(100)$ (see Table \ref{table_one}). Thus, the simulation best describes an equilibrated, topographically steered flow with significant along-stream variability. At the end of this section, we will address the rest of the simulations, when we present the topographic PV fluxes just derived in the previous section.

The mean circulation is characterized by a bottom-intensified anticyclonic boundary current associated with the southward facing (decreasing $y$ direction) sloping bottom localized near the western wall ($x=0$), and a cyclonic boundary current associated with the northward facing sloping bottom localized near the eastern wall ($x=250$km). Atop the ridge the mean flow is bottom intensified with a vertical expression that reaches up to the surface, and follows along f/H contours just north of the ridge crest.

We examine the vertical structure of the mean flow along the path of (time) mean transport streamlines that are calculated by inverting the 2D Laplacian of the (barotropic) vorticity ($\psi=\nabla^2\overline{\Omega}$, where $\Omega=\hat{\mathbf{k}}\cdot\nabla\times\langle\mathbf{u}_{h}\rangle$ and brackets represent vertical integration). While in the presence of sloping topography there is no longer a barotropic mode but instead a bottom intensified topographic Rossby wave \citep{rhines1970edge}, the depth scale of topographic Rossby waves associated with the mean (boundary current) circulation atop the ridge is relatively of the same order than the depth at the ridge crest $H_{p}\sim O(500m)$. Moreover, given that atop the ridge, all shear and baroclinicity is concentrated towards the bottom, the vertically integrated flow fields represent a good approximation of the dominant (throughflow scale) dynamics that are associated with topographically locked Rossby waves. The transport streamlines, thus, represent a good measure of the throughflow pathway.

\begin{figure}[t]
\centerline{\includegraphics[width=295pt]{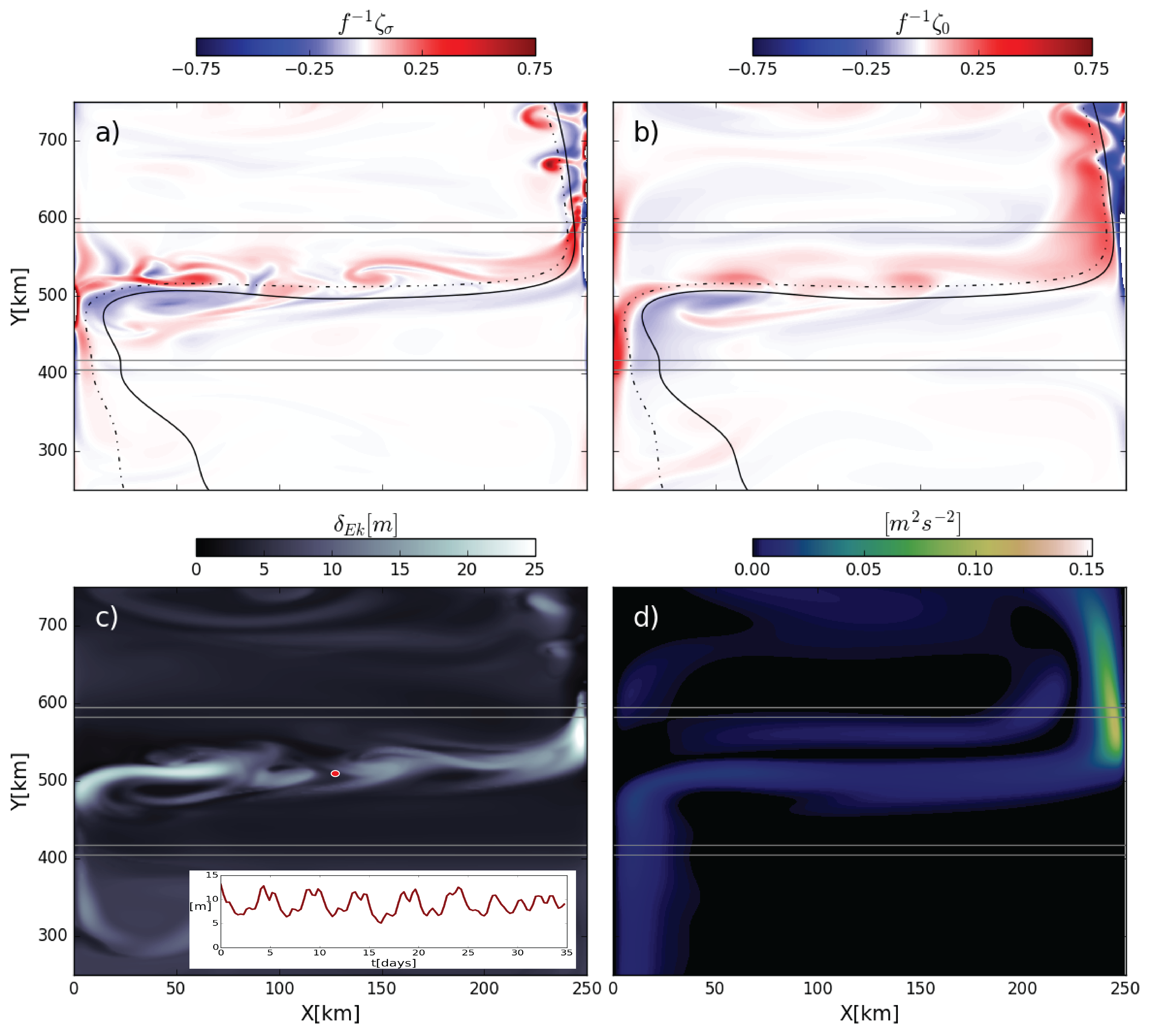}}
\caption{Snapshots of relative vorticity within the bottom boundary layer $\zeta_{\sigma}$ (a) and at the surface $\zeta_{0}$ (b), both scaled by the Coriolis parameter, in simulation $Cst_{ns}$. The span of the bottom ridge centered at y=500km is shown by the light gray contours, with the mean streamlines $\psi_{l}=0.5Sv$ (dashed-dot black) and $\psi_{c}=1.5$Sv (solid black) representing the mean transport pathway. Also shown is the bottom boundary layer thickness $\delta_{Ek}=\sqrt{2A_{\nu}/f}$  (c), which displays significant along-ridge variation due to (bottom intensified) baroclinic eddies. A time series of $\delta_{Ek}$ along the path of the mean flow away from boundary currents (red circle in c) is also shown. Lastly, the surface mean kinetic energy $(1/2)\overline{\mathbf{u}}_{h}^2$ is shown in (d), with a maximum along the cyclonic boundary current.\label{fig:zero}}
\end{figure}

To better understand the changes experienced by the throughflow as it navigates topography, we examine flow quantities along individual streamlines (e.g. $\psi=\{0.5,1.5, 2.5\}\;Sv$), recognizing that streamlines can diverge from each other and approximate the path of the throughflow (Fig. \ref{fig:zero}a-b). 

Along the path of the mean flow atop the ridge crest, we find a laterally and temporally varying vertical (turbulent) viscosity $A_{\nu}$ as a result of along-stream changes on the bottom intensified current, and eddies that separate from the western boundary current that propagate along $f/H$ contours. As a result, the thickness of the bottom boundary layer $\delta_{Ek}=\sqrt{(2A_{\nu}/f)}$ varies along the path of the flow atop the ridge, with the greatest thickness within the vicinity of the boundary currents ($\delta_{Ek}\approx25m$). Along the $f/H$ path of the bottom-intensified flow away from boundary currents, $\delta_{Ek}\approx 10-12m$ (Fig. \ref{fig:zero}c). Furthermore, baroclinic  bottom intensified eddies promote thickness changes to the boundary layer $\delta'_{Ek}\pm3m$ off the mean value (see Fig. \ref{fig:zero}c, inplot), with a timescale of approximately 4-5 days. The ratio of frequency on boundary layer thickness variability to inertial frequency $r_{\omega}=\omega_{bbl}/f \ll1$, which means that eddies allow the bottom intensified flow along $f/H$ contours to escape buoyancy shutdown \citep{ruan2016bottom}. Moreover, since $\mathcal{T}_{shut}>>\gg \mathcal{T}_{inertial}$, frictional spindown along the ridge is the dominant process, and with the bottom boundary layer being able to escape buoyancy shutdown associated with the mean flow.

\begin{figure}[t]
\centerline{\includegraphics[width=15cm,height=9cm]{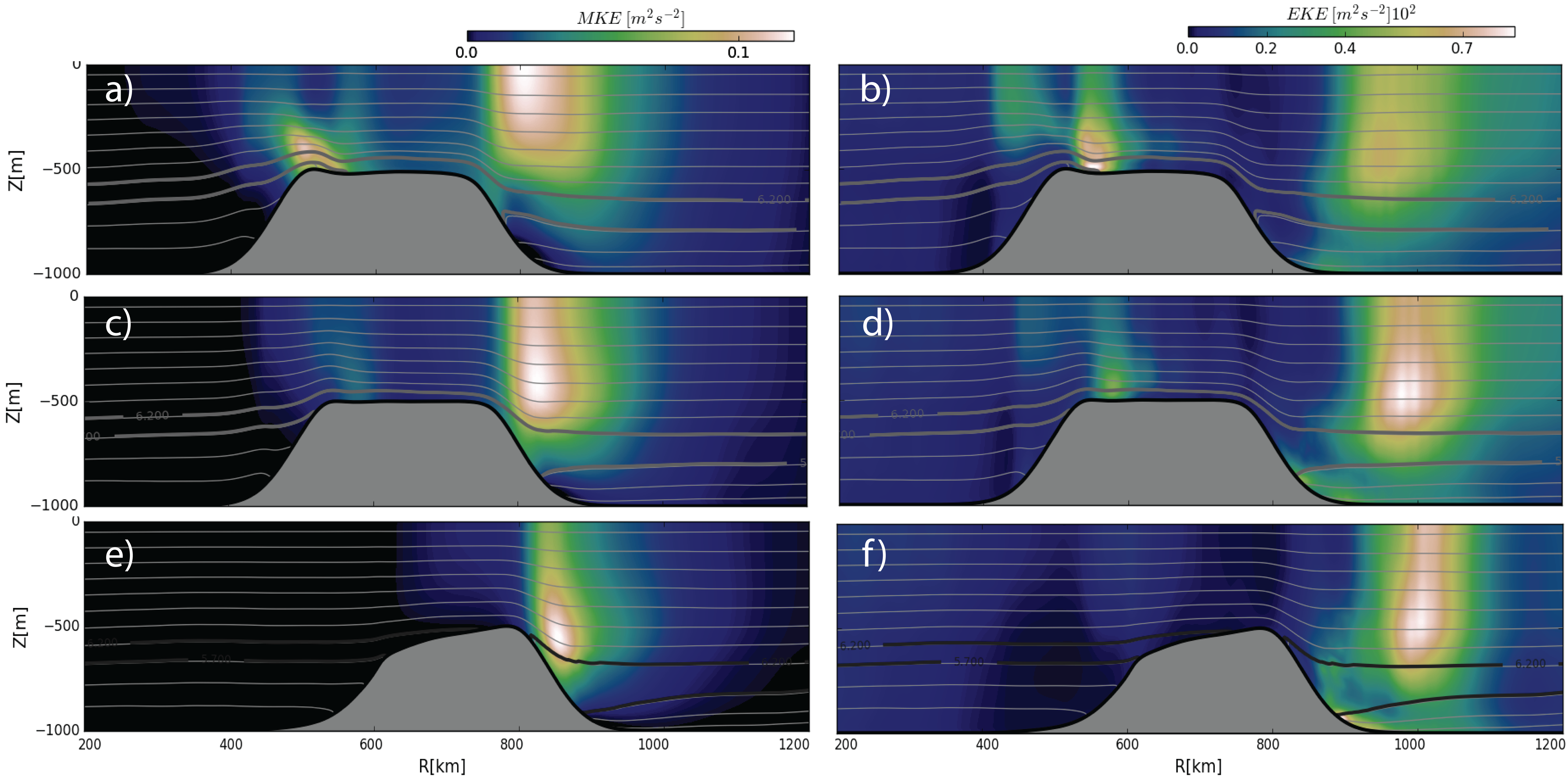}}
\caption{Along stream mean kinetic energy (MKE) (left column) and eddy kinetic energy (EKE) (right column) at streamlines $\psi_{l}=0.5Sv$ (a, b), $\psi_{c}=1.5Sv$ (c,d) and $\psi_{u}=2.5Sv$ (e, f). Light gray contours represents isotherms, plotted at every $0.5^\circ\:C$. The streamline $\psi=2.5Sv$ crosses the ridge east of the crossing of the other two streamlines. The three streamlines do, however, align and exhibit a similar behavior downstream from the ridge crest. \label{fig:one}}
\end{figure}

Both the mean and eddy kinetic energy (MKE and EKE respectively) along the western boundary current are bottom intensified (Fig. \ref{fig:one}a-d, at $R\approx550$km). Their maximum value are located close to each other, and decay rapidly as the flow moves along the ridge crest. MKE and EKE associated with the cyclonic boundary current are laterally apart from each other ($\sim 150$km), with the EKE maximum taking place downstream (EKE is maximum at R$\approx1000$km downstream from the ridge, see Fig.  \ref{fig:one}b,d;  MKE maximum at $R\approx850$km, see Fig. \ref{fig:one}a,c). 

The vertical structure of MKE and EKE along the cyclonic boundary current is suspect (\textit{e.g.} MKE behavior seen at R=850km in Fig. \ref{fig:one}c, and EKE behavior at R=1000km in Fig. \ref{fig:one}d). Both are middepth intensified, above the isotherm $T=6.2^\circ\:C$ (Fig.\ref{fig:one}a-d). Their vertical structure decays rapidly at increasing depths and slowly towards the surface. This suggests that both MKE and EKE retain the vertical structure of a (bottom intensified) topographically Rossby wave, a behavior expected for MKE since it lies above sloping bottom but not for EKE which lies downstream, away from sloping topography. 

\begin{figure}[t]
\centerline{\includegraphics[width=325pt]{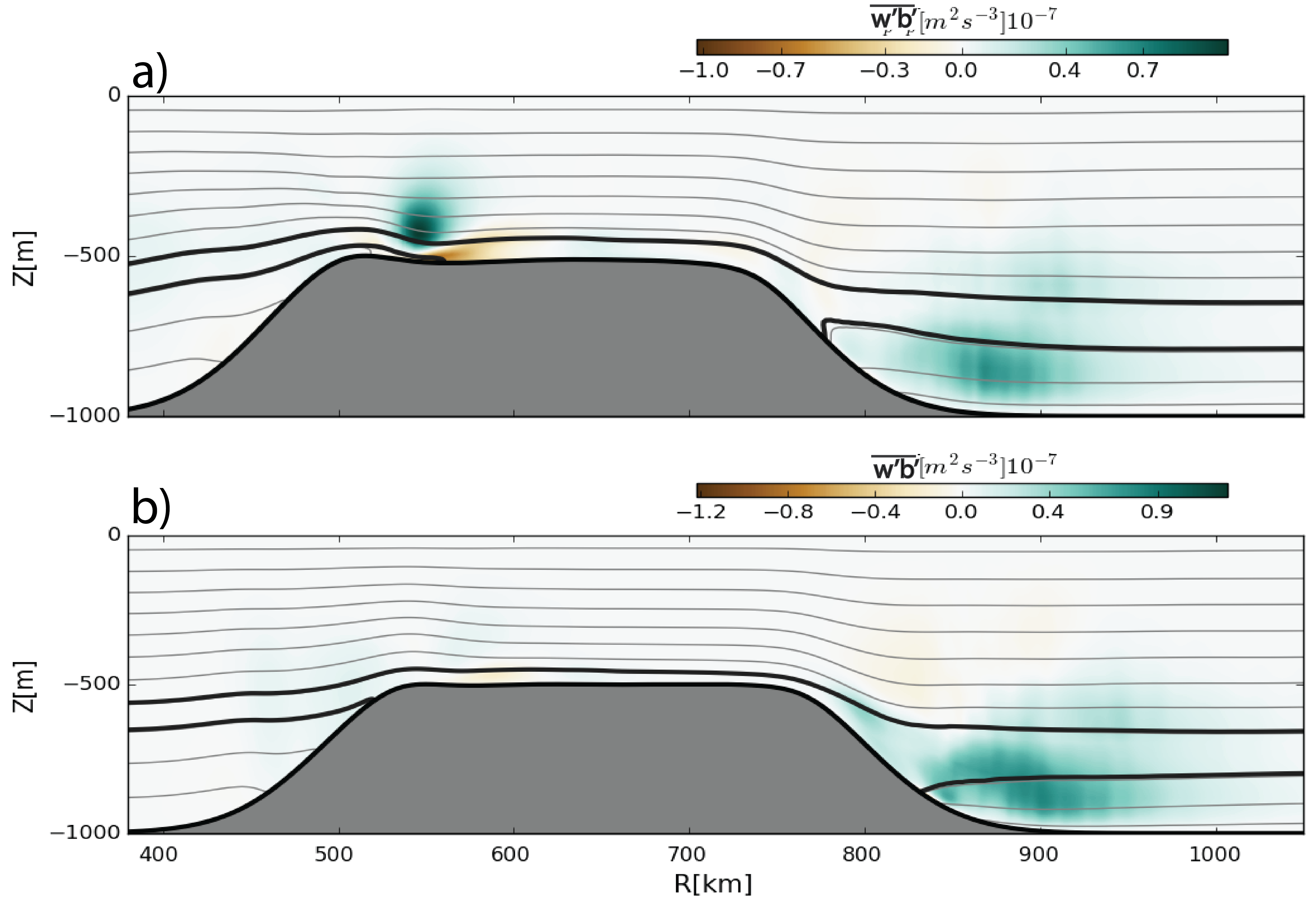}}
\caption{Buoyancy production ($\overline{w'b'}$) along the $\psi=0.5\;Sv$ (a) and $\psi=1.5\;Sv$ (b) time mean streamlines. Temperature is plotted at every $0.5^\circ\;C$ (grey contours). Thick black contours represent the isotherms
$\overline{T}_{1}=5.7^\circ\;C$ and $\overline{T}_{2}=6.2^\circ\;C$, which define the layer with anomalous low PV. At distance $R=550$km along the path of the streamlines, buoyancy production $\overline{w'b'}>0$ associate with the anticyclonic (western) boundary current can be seen clearly along $\psi=0.5Sv$, and only faintly along $\psi=1.5$Sv, due to bottom intensified vertical shear that introduces a vertical tilt in the velocity structure. For both streamlines, buoyancy production around $R=870$km can be seen at intermediate depths ($z=-800$m), suggesting a process of instability that is localized to near bottom isotherms that tilt rapidly in the along-slope direction.\label{fig:two}}
\end{figure}

The along stream behavior of buoyancy production $\overline{w'b'}$, associated with the conversion of available potential energy into eddy kinetic energy by baroclinic instability, provides further insight into the vertical structure of the mean and eddying fields. Associated with the western intensified boundary current (Fig. \ref{fig:two}a, at R=550km) there is a local maxima in buoyancy production, a sign of (baroclinic) instability growth associated with the bottom intensified boundary current.

Buoyancy production shows another local maxima at intermediate depths, associated with the instability of the cyclonic boundary current ($R\approx900$km in Fig. \ref{fig:two}b). Such local maxima is located below the local EKE maxima of the cyclonic boundary current and 100 km upstream (Fig. \ref{fig:one}d at $R=1000$ km). This suggests that eddies on the cyclonic boundary current are advected (or self advect) roughly 100km before these reach finite amplitude. Furthermore, buoyancy production implies growth by baroclinic instability, a mechanism that requires the \textit{relaxation} of tilting buoyancy/temperature surfaces in the across-stream direction and, thus, the presence of a cross-stream PV gradient.

Lateral PV gradients can be approximated by cross-stream layer thickness gradients and so we focus on an intermediate layer defined by two isotherms:  $T_{c}=\{5.7, 6.2\}^\circ\;C$ (thick black contours in Fig. \ref{fig:two}a,b). At first glance, there appears to be a cross-stream thickness gradient by comparing the behavior of the isotherms in Fig. \ref{fig:two}a,b at $R=800$km along two spatially separated streamlines: $\psi=1.5Sv$ (Fig. \ref{fig:two}b) and $\psi=0.5$Sv (Fig. \ref{fig:two}a). 

\begin{figure}[t]
\centerline{\includegraphics[width=300pt]{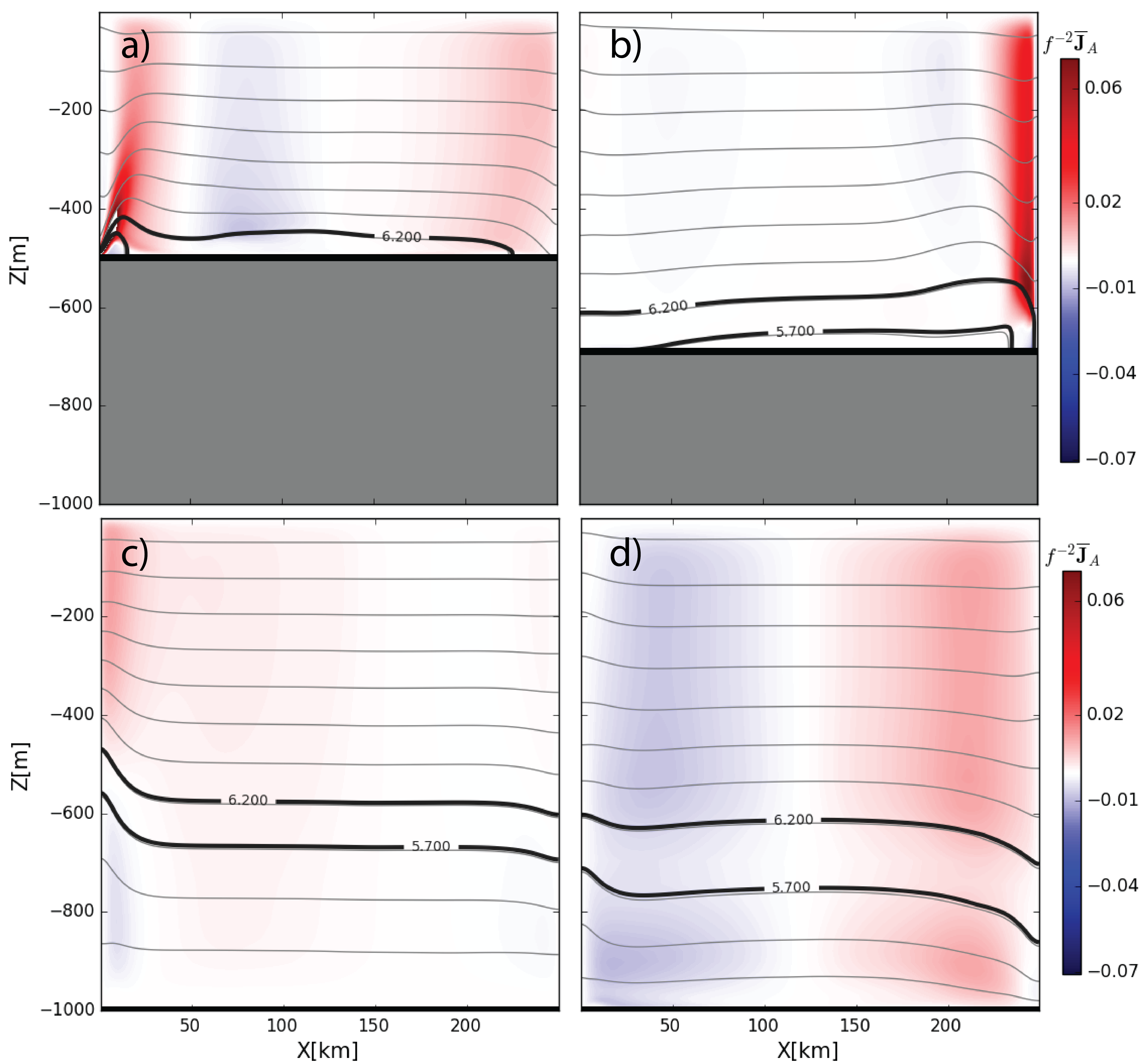}}
\caption{Zonal (x,z plane) sections of time-mean, northward advective potential vorticity flux ($\overline{vq}$) along the ridge crest (a) where the mean flow associated with the boundary current is strongest, at $y=550$km where the northward flow associated with the cyclonic boundary current is strongest (b). In addition, section upstream at y=200km (c) and another downstream 800 km (d) from the ridge crest are shown. The location of the mean isotherms $\overline{T}_{1}=5.7^\circ\;C$ and $\overline{T}_{2}=6.2^\circ\;C$, are shown in thick black contours. The isotherms incrop towards the sloping bottom in the presence of the cyclonic boundary current (c,d), resulting in a layer with vanishing vertical stratification and vanishing potential vorticity flux. \label{fig:three}}
\end{figure}

Zonal sections of the mean northward advective PV flux $\overline{\mathbf{J}}_{A}\cdot\mathbf{j}$ further reveal the cross-stream behavior of isotherms. We focus on 4 locations: a section along the ridge crest at $y=500$km where the local maxima of $\overline{\mathbf{J}}_{A}$ is associated with the bottom intensified anticyclonic boundary current (Fig.\ref{fig:three}a), a section at $y=550$km north of the ridge crest where the local maxima of $\overline{\mathbf{J}}_{A}$ is associated with the bottom intensified cyclonic boundary current (Fig. \ref{fig:three}b); another section upstream from the ridge at $y=200$km (Fig. \ref{fig:three}c) and lastly a section at $y=800$km far downstream from the ridge (Fig. \ref{fig:three}d).

At the section along $y=550$km, the isotherms $T_{c}=\{5.7^\circ\:C, 6.2^\circ\:C\}$ incrop towards the sloping bottom topography, with the advective PV going to zero within the space (area) bounded by these isotherms (z=600m, x=240km in fig. \ref{fig:three}b). The mean advective PV can be approximated as $\mathbf{\overline{J}_{A}}\approx\overline{v}\;\overline{q}\mathbf{\hat{j}}$, since baroclinic growth takes place farther downstream. The vanishing of the mean advective PV is then associated with a vanishing mean (Ertel) potential vorticity $\overline{q}\approx0$, given the observed isothermal tilts (thus vanishing stratification), and $\overline{v}\neq0$, even within the bottom boundary layer (Fig. \ref{fig:one}a, c, e). 

The stratification associated with the incropping layer defined by isotherms $T_{c}=\{5.7^\circ\:C, 6.2^\circ\:C\}$ (Fig. \ref{fig:three}b) resembles that of a mixed layer front, associated with the (cross-slope) boundary current flow. This is a different configuration from idealized quasi 2D bottom mixed layer fronts in the literature. Bottom mixed layer fronts within the bottom boundary layer with an orientation like the one observed in Fig. \ref{fig:three}b must then be associated with a net along-slope (westward) Ekman driven transport of buoyancy/temperature anomalies, i.e. away from the eastern wall. Such Ekman transport of warm water anomalies then promotes a convectively unstable stratification, hence the formation of an incropping bottom mixed layer front. 

\begin{figure}[t]
\centerline{\includegraphics[width=350pt]{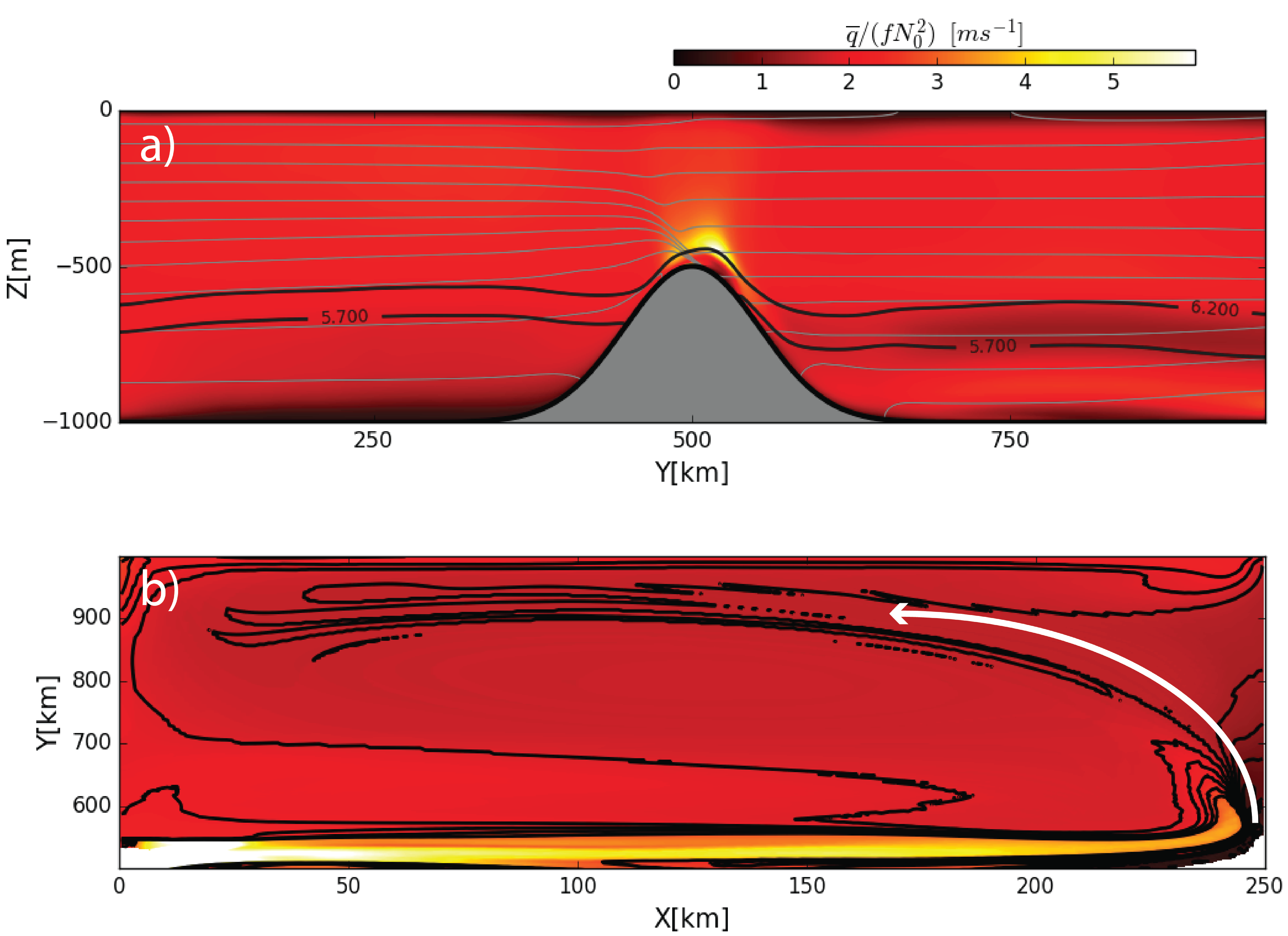}}
\caption{a) Cross-ridge spatial distribution of mean potential vorticity $\overline{q}$ at $x=125$km, away from boundary currents, showing a low PV tongue north of the ridge crest, between isotherms $T_{1}=5.7$ and $T_{2}=6.2$ (thick black contours). Grey contours represent isotherms plotted every $0.5^\circ$C, evaluated at the west wall $x=0$, for comparison (they show the location of the bottom front, where the local maximum PV anomaly is located). b) potential vorticity along isentrope $T_{c}=6.0^\circ\;C$ north of the ridge crest ($y>500$km). This isentrope is located within the layer with low PV anomaly highlighted in a). Black contours now represent PV contours along the isentrope, showing the spread of low PV tongue associated with the mean cyclonic boundary current.\label{fig:four}}
\end{figure}

Downstream from the ridge, the layer defined by the isotherms $T_{c}=\{5.7^\circ\:C, 6.2^\circ\:C\}$ shows a vanishing advective PV flow (Fig. \ref{fig:three}d). Given the strong barotropic nature of the basin-scale circulation downstream from the ridge and that $\overline{J}_{A}$ does not change sign in the vertical, these imply a vanishing of the potential vorticity $q\approx0$ within the layer defined by the $T_{c}$-isotherms. Our argument is supported by the spatial distribution of mean potential vorticity across the ridge at the zonal section $x=125km$ (Fig. \ref{fig:four}a). Downstream from the ridge, the layer defined by isotherms $T_{c}=\{5.7^\circ\:C, 6.2^\circ\:C\}$ shows a near zero PV anomaly (when compared to the PV values above of below). Given that PV cannot be changed within an isentropic surface \citep{haynes1987evolution,haynes1990conservation}, the observed PV distribution must then be associated with non-local advective transport. 

\begin{figure}[t]
\centerline{\includegraphics[width=325pt]{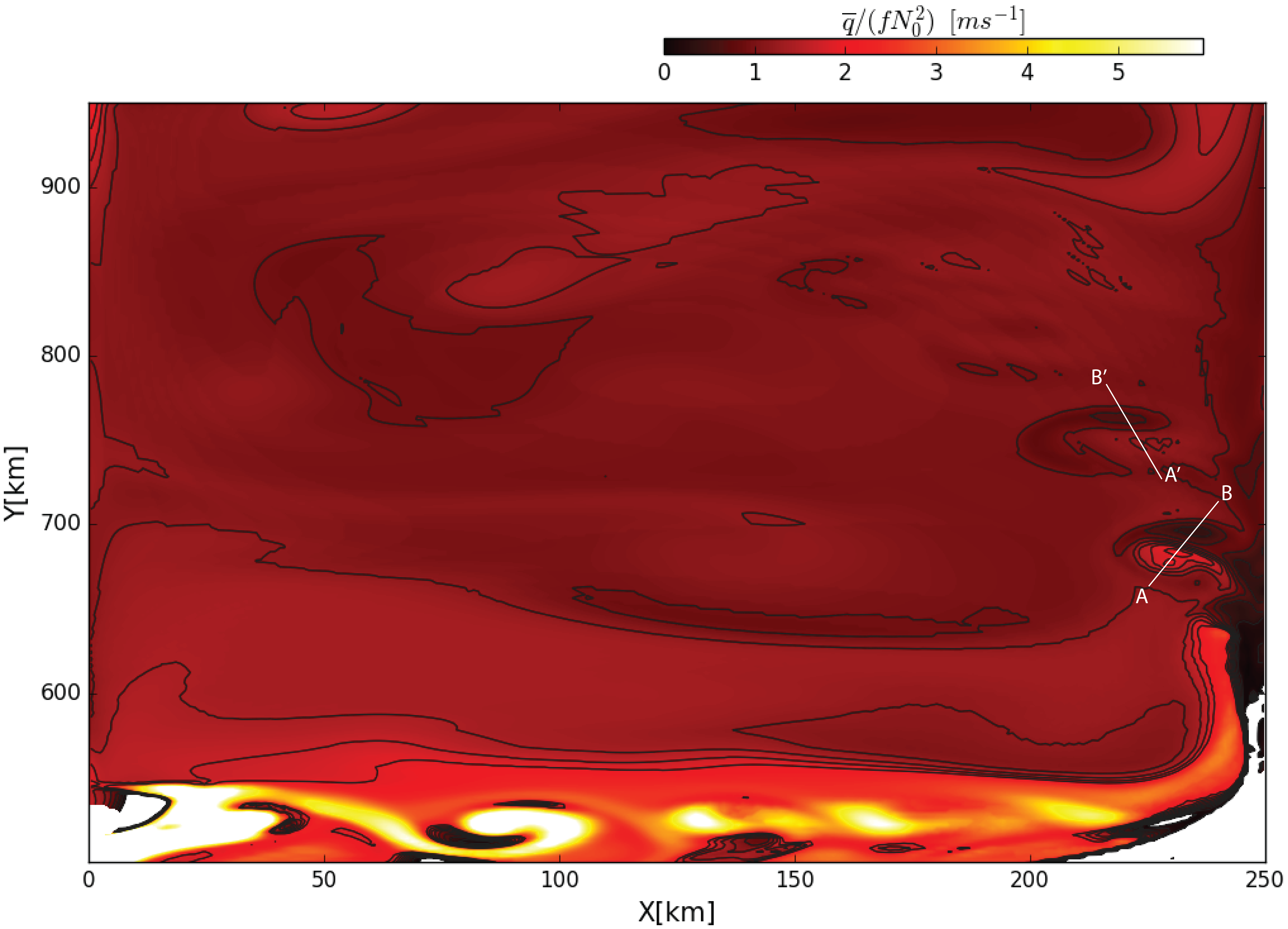}}
\caption{Snapshot of isentropic Ertel PV in simulation $Cst_{ns}$ ($T_{c}=6.0^\circ\;C$) north of the ridge crest ($y>500$km), located within the layer with low PV anomaly highlighted in Fig. \ref{fig:four}a. Black contours represent PV levels $\overline{q}/fN_{0}^2=1.5$. Dipoles are continuously generated from the region of baroclinic instability growth $(240km,600km)$, and advect low PV anomalies (anticyclonic vorticity) into the interior. A dipole can be seen at $(x,y)=(225km,680km)$, with the line $\overline{AB}$ bisecting it. Dipole rotate cyclonically as they are advected by the mean flow, as shown by a second dipole downstream $(x,y)=(215km,750km)$, where the line $\overline{A'B'}$ shows the respective orientation. Cyclonic (high PV) anomalies can be seen propagating along the ridge near the crest ($y=550$km), dissipating near the cyclonic boundary current ($x\approx 220km$).\label{fig:five}}
\end{figure}

We now consider the isentropic PV  (IPV) distribution along the $T_{0}=6.0^\circ\:C$ surface. The IPV shows a spread of low PV associated with the signature of bottom mixed layer frontal watermass into the interior (beginning at x=240km, y=500km in Fig. \ref{fig:four}b), roughly in the direction of the mean flow (white arrow). A snapshot of IPV along the same $T_{0}=6^\circ\:C$ surface for a day near the end of our (7 year) simulation shows eddies (dipoles) advecting low PV anomalies within their anticyclonic core, into the interior, away from their formation site (Fig. \ref{fig:five}, at $x=225$km, $y=680$km). These eddies appear to be advected downstream by the mean flow, as well as rotate cyclonically, a sign of an unbalanced dipole with one core strong than the other \citep{manucharyan2013generation}. As the dipoles reach finite amplitude, these appear to get dissipated potentially by the strong shear of the mean (separating) boundary current. 

IPV maps also show the advection of anomalous high PV along the ridge crest (Fig. \ref{fig:four}, and Fig. \ref{fig:five}). Advection of high PV anomalies is associated with the bottom intensified baroclinic eddies, predominantly cyclonic (e.g. Fig.\ref{fig:five} $x=90$km, $y=550$km) and along $f/H$ contours.

\subsection{Topographic PV Fluxes}

The sources of PV anomalies shown in Figs. \ref{fig:four} and \ref{fig:five} are associated with the interaction of the bottom-intensified flow with the frictional bottom that result in local intersection of temperature/buoyancy surfaces with the sloping bottom in regions of strong curvature of the flow as it navigates the ridge. In this section we complement the IPV and along-streamline analysis of the vertical structure of the mean and eddy fields, by presenting an (Eulerian) volume integrated PV budget (see Appendix \ref{IntegralPV} for a detailed description). Our previous analysis suggests net topographic PV fluxes to play an important role balancing the budget as these are associated with PV sources.

\begin{figure}[t]
\centerline{\includegraphics[width=375pt]{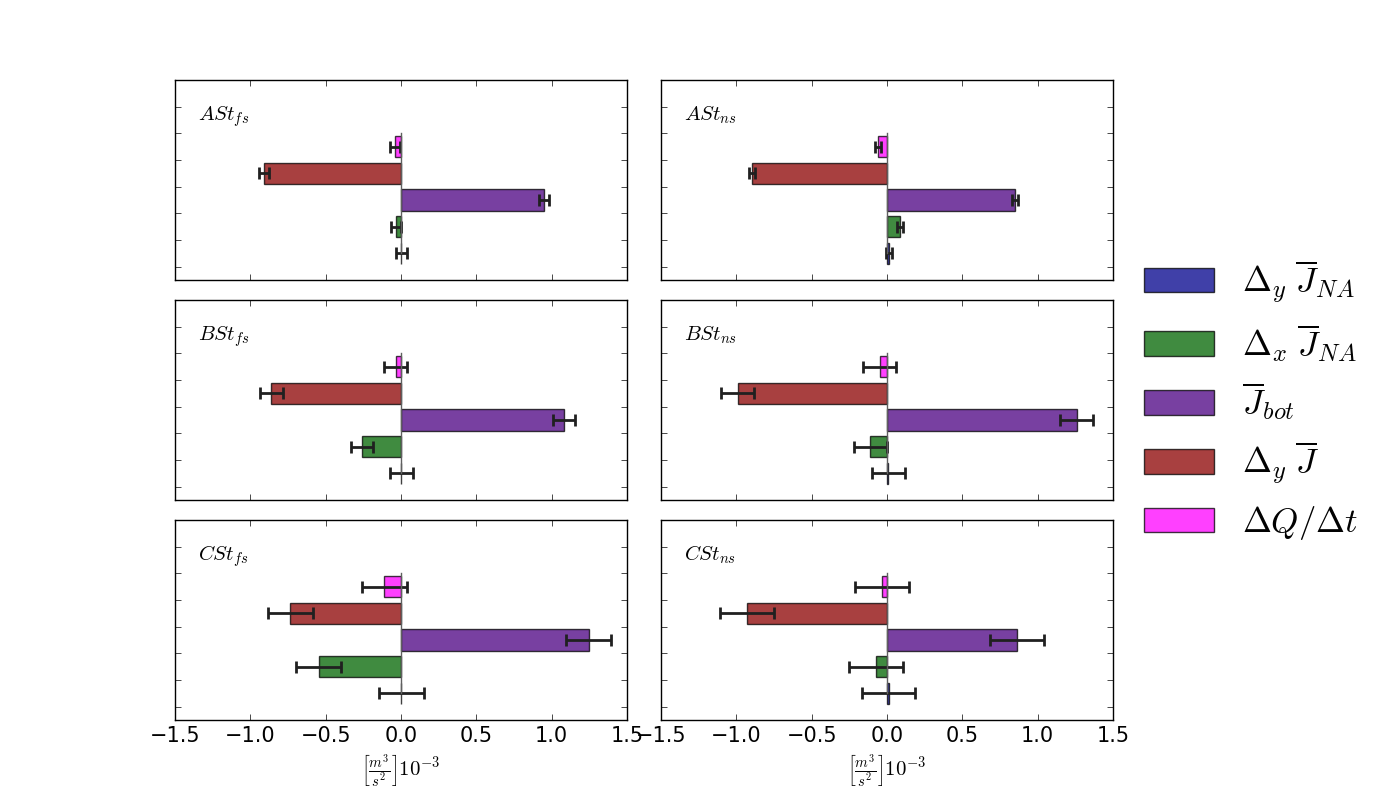}}
\caption{Integral PV balance (\ref{IntConstrain1}) for all simulations. The error bars denote the residual error ($\pm\Delta$) in the balance. Note that the net northward dissipative PV flux ($\Delta_{y}\overline{J}_{NA}$) is vanishingly small in all simulation.\label{fig:six}}
\end{figure}

The control volume considered here is delimited by one cross channel section far upstream and another downstream from the ridge (at $y=100$km and $y=900$km). We calculate the volume integrated balance of PV in (\ref{PV_eqn}), over a sufficiently long time that the LHS of (\ref{PV_eqn}) is vanishingly small (Fig. \ref{fig:six}), and the evolution of PV is driven by boundary sources and interior redistribution (\textit{e.g.} mean flow or eddy advection).

In all simulations the net northward PV flux is negative $\Delta_{y}(\mathbf{\overline{J}}\cdot\mathbf{\hat{j}})<0$, implying a larger (northward) PV flux entering the control volume (upstream from the ridge), 
where $\Delta_{y}(\cdot)=(\cdot)_{y=900}-(\cdot)_{y=100}$ represents the difference in the integrated fluxes down and upstream. In addition, the dominant contribution to the net northward flux is advective \textit{i.e.} $\Delta_{y}(\overline{\mathbf{J}}\cdot\mathbf{\hat{j}}) \approx \Delta_{y}(\mathbf{\overline{J}}_{A}\cdot\mathbf{\hat{j}}) < 0$ (Fig. \ref{fig:six}). 

The net loss of northward PV flux is approximately balanced by a net positive topographic PV flux ($\overline{J}_{bot}>0$), in all simulations (Fig. \ref{fig:six}) \footnote{ROMS does not explicitly solve the integral PV equation (\ref{IntConstrain1}) at every time-step. As a consequence, we don't expect that our integral PV budget will close exactly, but we find relatively good agreement.}. Thus

\begin{equation}\label{approx_bal}
\Delta_{y}\overline{J}_{A} \approx -\overline{J}_{bot}
\end{equation}

This approximate balance is consistent with our previous findings, where the high PV anomalies remain atop the ridge, whereas the low PV anomalies get advected downstream and homogenized, thus reducing the value of the background PV downstream from the ridge, and so reducing the northward advective PV flux downstream out of the control volume.

\begin{figure}[t]
\centerline{\includegraphics[width=15cm,height=10cm]{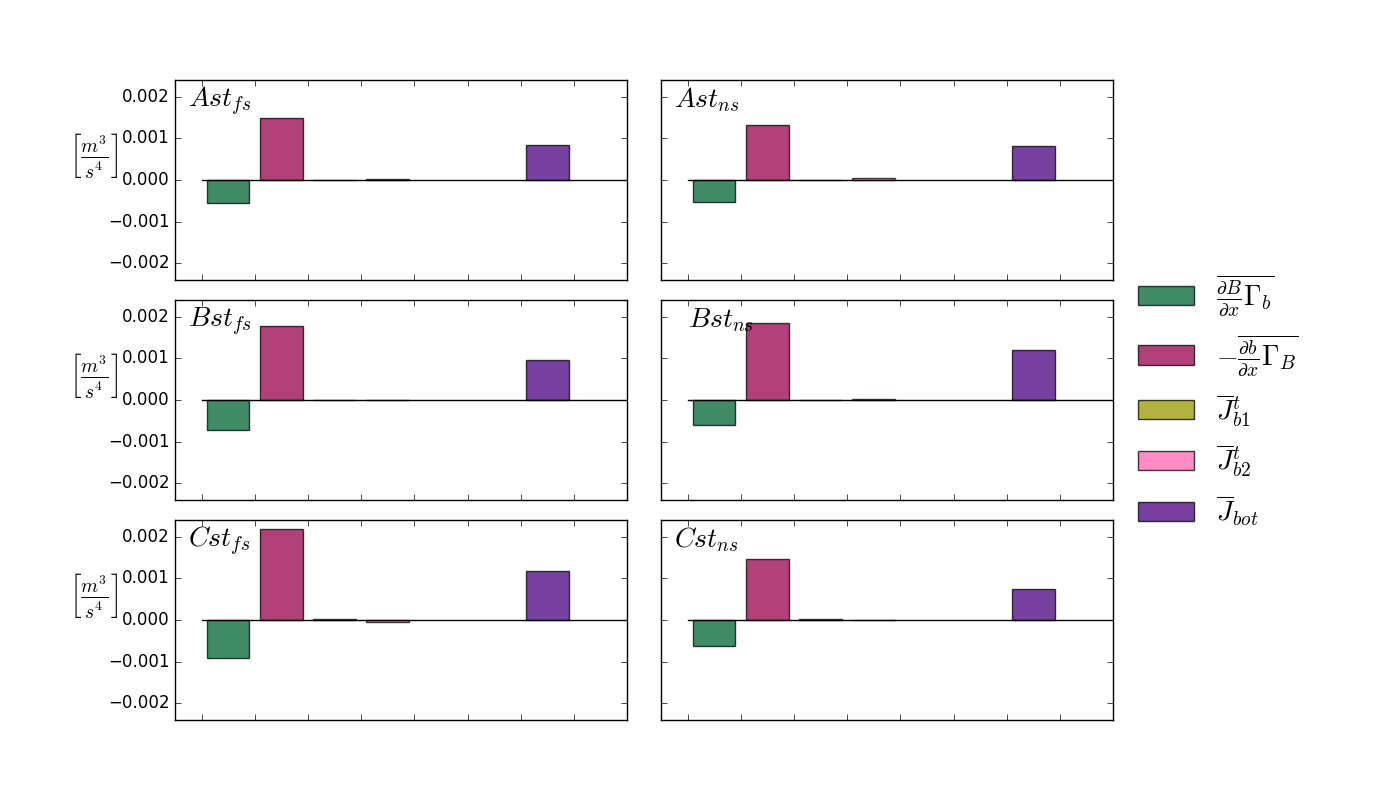}}
\caption{Net contributions of the terms in the decomposition of the net topographic potential vorticity flux $\overline{J}_{bot}= \overline{J}^t_{bot} + \overline{J}^s_{bot}$, where $\overline{J}^t_{bot}=\overline{J}^t_{b1}+\overline{J}^t_{b2}$ and $\overline{J}^s_{bot}=[\overline{B_{x}\Gamma_{b}}-\overline{b_{x}\Gamma_{B}}]h_{y}$ In all simulations, $\overline{J}^t_{bot}\approx 0$, making a vanishingly small contribution to the integral PV balance. \label{fig:seven}}
\end{figure}

\begin{figure}[t]
\centerline{\includegraphics[width=475pt]{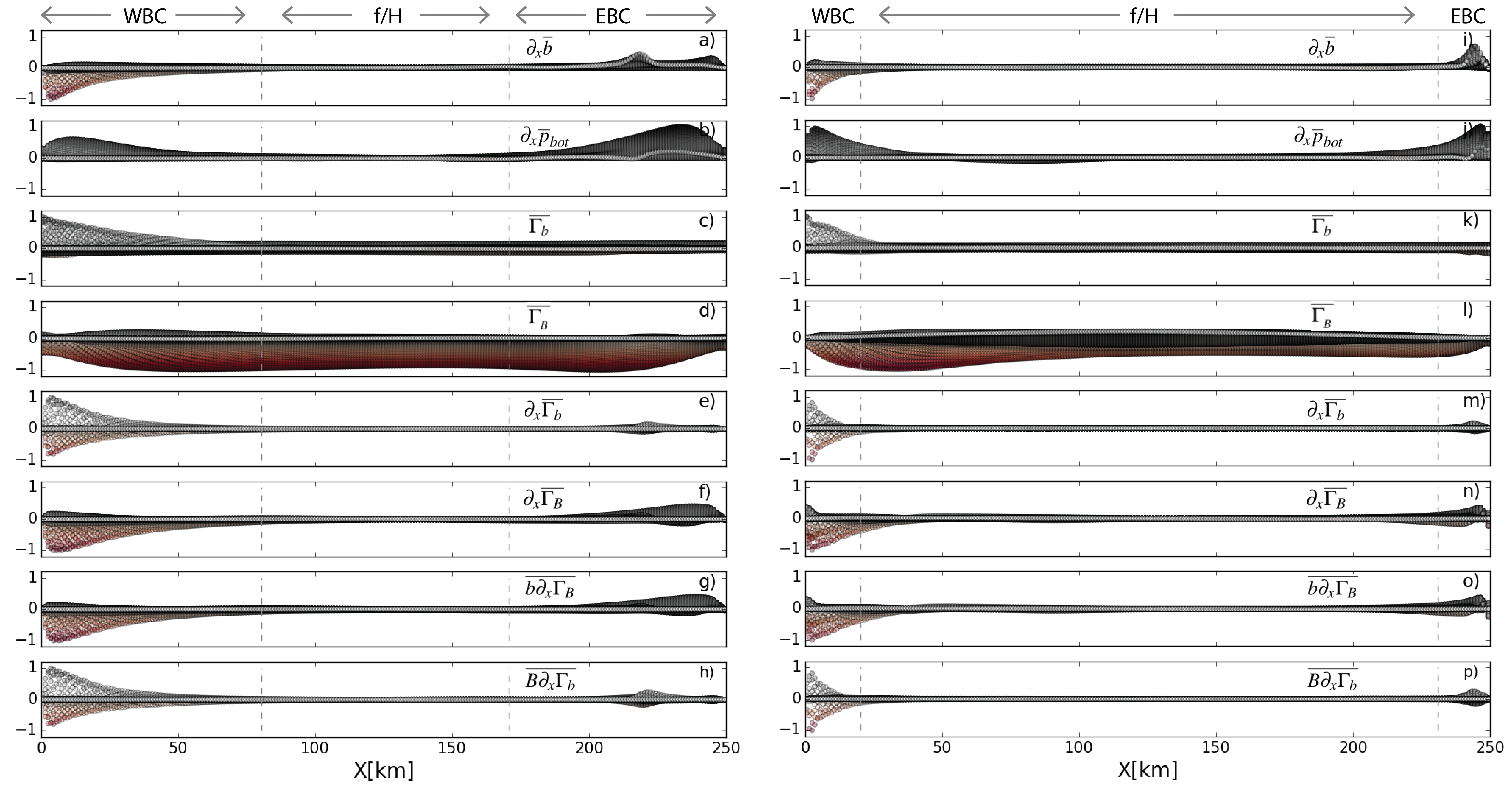}}
\caption{Along-ridge spatial variation of the terms associated with the second term in (\ref{sJbot4n}), a term in the decomposition of the topographic PV flux associated with the along $f/H$ contributions (Left panels from simulations $Ast_{ns}$ and right panel the terms from simulation $Cst_{ns}$). Denoted above the panels, shown by dashed vertical lines, are the regions of influence from western and eastern boundary currents (WBC and EBC), as well as the area where the flow is along f/H contours. All terms are normalized, and evaluated at 5m above the bottom, within the bottom boundary layer, and only taking into account values spanning the extent of the ridge ($400\;km<y<600\;km$). \label{fig:eight}}
\end{figure}

Following (\ref{two_decomp}) and (\ref{sJbot2n}), we now look at the decomposition of the topographic PV flux term among the different expressions. We find that $\overline{J}^t_{bot}$ makes no contribution to the net flux in all simulations, so that $\overline{J}_{bot}\approx\overline{J}^s_{bot}$ (Fig. \ref{fig:seven}). Furthermore, we use the decomposition in (\ref{sJbot4n}), where we separate the terms (\ref{sJbot2n}) between those evaluated at the walls and those along $f/H$ contours away from the influence of the boundary currents (see Fig. \ref{fig:eight}, which shows the terms that make the dominant PV terms in (\ref{sJbot2n})). Such decomposition allows to focus on the spatial distribution of the (local) fluxes on three regions: one associated with each lateral boundary current and one associated with the flow along $f/H$ contours, near the ridge crest. The dynamics associated with the mean flow along $f/H$ contours best resembles idealized quasi-2D studies of bottom boundary layers at a slope \citep{maccready1991buoyant, benthuysen2012friction}. 

In all simulations, we find that the greatest contribution to the net topographic PV fluxes takes place within the region of influence of lateral boundary currents, where the mean flow departs greatly from f/H contours (Fig. \ref{fig:eight} m-p). All variables except $\overline{\Gamma}_{B}dh/dy$, decay drastically away from the influence of lateral boundary currents (Fig. \ref{fig:eight},l). This can be associated with the fact that the temperature surfaces intersect the bottom in the vicinity of boundary currents associated with the bottom fronts (with high and low PV anomalies), but the along-slope flow experiences little of such behavior.

\begin{figure}[t]
\centerline{\includegraphics[width=400pt]{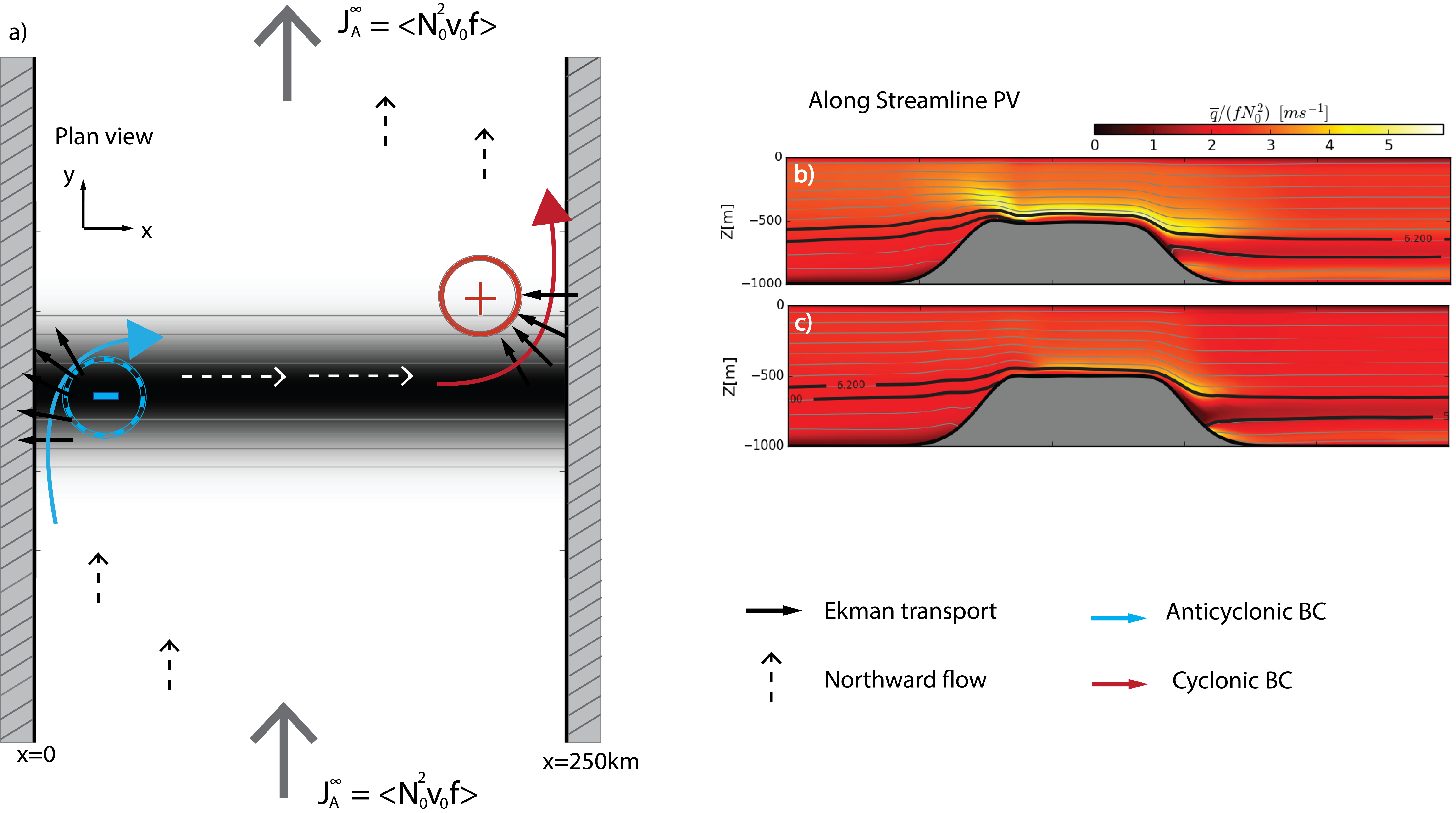}}
\caption{a) Plan view diagram of the time-mean circulation with arrows representing the various components of the flow field. White dashed arrows represent the along-slope flow atop the ridge, and are located slightly north of the crest. Also shown, PV distribution along the streamlines $\psi=0.5$Sv (b) and $\psi=1.5$Sv (c) capturing the locations of the PV maxima and minima associated with frictional Ekman transport.\label{fig:nine}}
\end{figure}

In the region associated with flow along $f/H$ contours, the mean flow (and thus the bottom boundary layer) is located near the ridge crest where the bottom slope approaches zero, resulting in a buoyancy shutdown timescale much greater than the stratified, frictional spindown timescale, \textit{i.e.} $\mathcal{T}_{shut}\gg\mathcal{T}_{spin}$ (see Table \ref{table_two}). This suggest that buoyancy effects within the bottom boundary layer are too slow to promote baroclinicity within the bottom boundary layer, and thus induce Ekman arrest. Thus, while we do observe downward Ekman transport within the bottom boundary layer (Fig. \ref{fig:nine}b), the vanishingly small contribution to the PV flux is likely associated with the location of the mean flow being close to the ridge crest, where the topographic slope almost vanishes, and eddy driven variability dominates. Another important factor to consider is the strong stratification upstream, associated with the high PV front, and the baroclinic instability that ensues, which continuously act to relax the temperature/buoyancy surfaces with respect to depth levels.

\section{Discussions and Conclusions}\label{d_and_c}

We have presented in this study an analysis of the potential vorticity budget and the vertical structure of a rotating, stratified throughflow across finite amplitude topography under channel geometry. We found that bottom boundary layer dynamics associated with boundary currents help promote the formation of \textit{high} PV anomalies along an anticyclonic western boundary current, and a \textit{low} PV anomaly associated with an eastern (\textit{pseudo} westward) cyclonic boundary current. The \textit{high} PV anomaly is advected along topographic contours atop the ridge. Meanwhile, the vanishingly \textit{low} PV anomaly, is advected downstream resulting in a mid-depth isentropic layer characterized by watermasses with anomalous low PV signature. 

The observed PV distribution associated with anticyclonic (net upslope transport) and the cyclonic boundary currents (net downslope transport) are, in a way, in accordance to the vertical stratification expected from basic PV principles of stretching and squashing associated with flow past topographic obstacles \citep{pedlosky2013geophysical, vallis2006atmospheric}. Such principles are based on linear, inviscid QG dynamics that assure PV conservation. Our work shows that when boundary layer dynamics are important, dissipative processes irreversibly amplify the expected stratification promoting strong high bottom stratification (a front and thus creating high PV anomaly) as the flow moves into the ridge crest, and a low bottom stratification (the mixed layer front with low PV). As eddies help redistribute PV anomalies non-locally, dissipative and nonlinear advective processes not typically included in first principles of PV conservation, are responsible for the net decrease of PV fluxes carried by the net northward transport. That is, in the absence of dissipative PV processes and eddies, one would expect a net zero northward PV flux as the flow navigates across finite amplitude topography.

Throughout our simulations, we found that buoyancy shutdown (thus Ekman arrest) is not a dominant process within the bottom boundary layer. This can be attributed to the fact that the along-slope flow takes place very close to the ridge crest where the slope Burger number approaches zero, effectively increasing the timescale required for buoyancy shutdown to take place. The location of the along-slope flow with respect to the ridge crest is largely determined by the location of the reversal of the background PV gradient, in this case given entirely by the topography slope. An increase in the background PV ($\beta$-plane dynamics) would lead to an along-slope flow further downstream, and thus an increased topographic slope. However, adding $\beta-$plane dynamics would also result in a domain scale western boundary current given our inflow-inflow conditions, like those in \citet{yang2000water}, largely altering the flow pattern. In reality, $\beta$-effects, baroclinic pressure forces (driven by contrasting density differences across topography), barotropic forcing (like the one considered here, due to the inflow and outflow) and surface forcing all play a role in determining the large scale patters as well as the local dynamical processes studied here. Our objective was to isolated the topographic effects on the throughflow.

Our choice of lateral walls has an influence on the orientation of the equilibrated bottom mixed layer front, which is perpendicular to isobaths and aligned with the mean (interior) flow. Nonetheless, imposing a step sloping bottom (coastal shelfs) instead of a lateral wall would still lead to net Ekman transport of warm water into the deep ocean in the direction perpendicular to a cyclonic boundary current. However, a step continental shelf break instead of a lateral wall could lead to a significant modification of the mean flow since such $f/H$ contours could provide topographic Rossby waveguides along the shelf, thus a separate pathway for the exchange of watermasses across basins, further complicating the model and analysis. The observed orientation of the bottom mixed layer was not explored by \citet{wenegrat2018submesoscale}, and the dynamics are likely very different due to the strong shear flow of the interior flow and the (rapid) changes experienced as it navigated the depth changes.

Our observed middepth eddy variability and subsequent eddy advection of low PV anomalies away from the middepth front, resembles the middepth boundary current of low potential vorticity studied in \citet{spall2008western}. There, eddies that grow from the middepth boundary current advect low PV waters into the interior. The major difference in our model is the significant mean flow advection into the interior, and the character of our flow: an equilibrated system. The equilibration of our throughflow allows us to quantify the net advective fluxes across the ridge that originate from the injection of vanishingly low PV waters from the bottom boundary layer, through a Eulerian PV volume budget.

We do not address the role of Kelvin waves and hydraulic control on PV sources, although the stratification pattern along the western wall (Fig. \ref{fig:nine}a) resembles that of a system experiencing hydraulic control. However, hydraulic control and thus the role of Kelvin waves in advective PV fronts, are largely transient problems in the absence of dissipation. Boundary currents require a viscous and frictional interaction with solid boundaries and, thus, it is unlikely that the observed stratification along the lateral walls is determined by inviscid wave processes.

A staple of hydraulic control theory is that PV along the ridge can exert a control on the upstream circulation \citep{whitehead1995critical, helfrich2003rotating}. Thus, it is possible that Kelvin waves, away from the influence of strong boundary current and bottom fronts, can propagate (advect) the PV anomalies found atop the ridge. In doing so, Kelvin waves would help drive the observed net northward PV flux by promoting a higher advective PV flux upstream from the ridge and a lower advective PV flux downstream. They would be doing so by advecting PV anomalies (strong/low stratification) along the walls, e.g. advect upstream a high PV front along the western wall and advectve a low PV front along the eastern wall downstream. Under such scenario, baroclinic eddies upstream and downstream from the ridge could potentially advect such anomalies from the walls into the interior, help raise the background PV value by a process of homogenization.

Based in our results its limitations, an interesting problem to which we can expand is that of an exchange in the presence of a destabilizing buoyancy forcing, for example, on a basin, resembling that by \citet{spall2010dynamics} which drives a convective overturning circulation across the ridge crest, while also retaining a large scale forcing (inflow/outflow) such that there is a barotropic (and baroclinic) pressure gradient across the ridge and thus a throughflow. Then, some particularly interesting questions arise in such scenario, for example how the boundary and interior distribution of PV anomalies may change by introducing such effects, and the presence of overflows. Such problem then represents the next step into generalizing our results.

%%%%%%%%%%%%%%%%%
%ACKNOWLEDGMENTS
%%%%%%%%%%%%%%%%%

\acknowledgments
This research was funded by NASA (NNX13AE28G, NNX13AH19G and NNX17AH56G) and the Mexican Council for Science and Technology (CONACyT following its abbreviation in spanish). We thank Charles Eriksen for helpful comments on an earlier draft, and Peter Rhines for many helpful discussions on the dynamics of flows across topography.

\appendix[A]\label{IntegralPV}
\appendixtitle{Integral PV Balance}

\subsection{Integral Constrains}

Consider a rectangular channel geometry with dimensions $(0\leq x\leq L)\times(0\leq y\leq M)\times(-h(y)\leq z\leq\eta)$, where $z=-h(y)$ is a finite-amplitude, meridionally isolated ridge, that is symmetric around $y=M/2$. A stratified flow across the ridge is driven by lateral boundary inflow/outflow that equivalent to a constant, large-scale surface pressure gradient across the ridge. We analyze the constrains the ridge has on the transport streamfunction and potential vorticity of the fluid. To better get a sense of the 3D character of the flow, we integrate the potential vorticity equation (\ref{PV_eqn}) over the channel domain, yielding an integral balance that incorporates both boundary sources and interior changes over a time interval. The time averaged integral equation is given by

\begin{eqnarray}\label{IntConstrain1}
\underbrace{\int_{A_{I}}\left[(\overline{\mathbf{J}}\cdot\hat{\mathbf{n}})\big\rvert_{_{y=0}}+(\overline{\mathbf{J}}\cdot\mathbf{\hat{n}})\big\rvert_{_{y=M}}\right]\;dA_{I}}_{\text{ Net Northward PV flux}} + \underbrace{\int_{A_{II}}\left[(\overline{\mathbf{J}}_{NA}\cdot\hat{\mathbf{n}})\big\rvert_{_{x=0}}+(\overline{\mathbf{J}}_{NA}\cdot\mathbf{\hat{n}})\big\rvert_{_{x=L}}\right]\;dA_{II}}_{\text{Net Dissipative PV flux}} \nonumber\\ 
= -\dfrac{\Delta Q}{\Delta t} -\underbrace{\int_{A_{bot}}\overline{\mathbf{J}}_{bot}dA_{bot}}_{\text{Net Topographic PV flux}}\nonumber\\
\end{eqnarray}

where the limits of integration on the left hand side are $A_{I}=(0\leq x\leq L)\times(-h\leq z\leq\eta)$ and $A_{II}=(0\leq y\leq M)\times(-h\leq z\leq\eta)$, the term $\Delta Q$ is the (mean) change of the volume integrated potential vorticity ($Q=\int_{V}qdV$) over time, and the integral term on the right hand side is taken over all variable bottom topography with area $A_{bot}$, where $\mathbf{J}_{bot}= \mathbf{J}\cdot\mathbf{\hat{n}}_{bot}$, with $\mathbf{\hat{n}}_{bot}$ the unit normal vector to bottom topography (outward-pointing). The integral balance in (\ref{IntConstrain1}) does not include surface PV fluxes due to the original assumption of no surface dissipative or diabatic forcing.

The time mean potential vorticity integral balance (\ref{IntConstrain1}) states that as the flow moves across finite amplitude topography, a net flux of PV due to lateral fluxes must be balanced by a net change of volume integrated PV, or a net flux of potential vorticity through bottom topography. Net lateral flux PV fluxes can take place as a result of a net northward flux (first integral term in \ref{IntConstrain1}), denoted from now on as $\Delta_{y}\overline{J}$, where $\overline{J}=\int_{A_{I}}\overline{\mathbf{J}}\cdot\mathbf{n}\;dA_{I}$ and $\Delta_{y}f=f(y=M)-f(y=0)$, or a net dissipative PV flux through lateral walls, the second integral term on left hand side of (\ref{IntConstrain1}), denoted as $\Delta_{x}\overline{J}_{NA}$. 

The dissipative potential vorticity flux normal to a lateral wall is independent of horizontal stratification in the primitive equations, despite $b_{y}\neq0$, resulting in the \textit{net dissipative lateral PV flux} given by

\begin{equation}\label{A2}
\Delta_{x}\overline{J}_{NA}=\int_{A_{II}}\left[\left(\overline{b_{z}\mathbf{F}^{y}} \right)\big\rvert_{_{x=L}} - (\overline{b_{z}\mathbf{F}^{y}})\big\rvert_{_{x=0}}\right]\;dA_{II}
\end{equation}

where $\mathbf{F}^{y}$ is the northward (y) component of the dissipation term $\mathbf{F}$. We decompose the integrands in (\ref{A2}) into a lateral harmonic term and a term resulting from the divergence of the frictional stress, in order to isolate the effect of lateral and vertical boundary walls \textit{i.e.} $\mathbf{F}=\mathbf{F}_{h} + \mathbf{F}_{v} = A_{_{H}}\nabla^2\mathbf{u}_{h} + \partial_{z}\left(A_{_{V}}\partial_{z}\mathbf{u}_{h}\right)$, where $A_{_{H}}$ and $A_{_{V}}$ are the coefficients of horizontal and vertical viscous momentum dissipation respectively. With this, the horizontal and vertical dissipative contributions to the integral terms in (\ref{A2}) are 

\begin{equation}\label{pal}
\int_{z=-h(y)}^{z=\eta}\overline{b_{z}\mathbf{F}^y_{h}}dz = H{<}\overline{N^2\mathbf{F}^y_{h}}{>} 
\end{equation}
and
\begin{equation}\label{kal}
\int_{z=-h(y)}^{z=\eta}\overline{b_{z}\mathbf{F}^y_{v}}dz = \rho_{0}^{-1}\left(\overline{N^2\tau^y}\right)\bigg\rvert^{z=0}_{z=-h(y)} -H{<}\overline{A_{v}v_{z}N^2_{z}}{>}
\end{equation}

where we have used $\tau^y_{b}=\rho_{0}(A_{v}v_{z})$ at top and bottom boundaries. With this, the \textit{net lateral dissipative flux of potential vorticity} is 

\begin{equation}\label{lat_int_terms}
\Delta_{x}\overline{J}_{NA} = \overline{J}_{NA_{1}} + \overline{J}_{NA_{2}}+\overline{J}_{NA_{3}}+\overline{J}_{NA_{4}}
\end{equation}
where
\begin{equation}\label{JNA1}
\overline{J}_{NA_{1}}=\int_{0}^{M}H\left[({<}\overline{N^2\mathbf{F}^y_{h}}{>})\big\rvert_{_{x=L}} - ({<}\overline{N^2\mathbf{F}^y_{h}}>)\big\rvert_{_{x=0}}\right] dy
\end{equation}
\begin{equation}\label{JNA2}
\overline{J}_{NA_{2}}=\int_{0}^{M}\left[\frac{(\overline{N^2\tau^y})\rvert_{_{x=L}}}{\rho_{0}} -\frac{(\overline{N^2\tau^y})\rvert_{_{x=0}}}{\rho_{0}}\right]_{z=0}\;dy
\end{equation}
\begin{equation}\label{JNA3}
\overline{J}_{NA_{3}}=-\int_{0}^{M}\left[\frac{(\overline{N^2\tau^y})\rvert_{_{x=L}}}{\rho_{0}} -\frac{(\overline{N^2\tau^y})\rvert_{_{x=0}}}{\rho_{0}}\right]_{z=-h}dy
\end{equation}
\begin{equation}\label{JNA4}
\overline{J}_{NA_{4}}=-\int_{0}^{M}H\left[<\overline{A_{v}v_{z}N^2_{z}}>\big\rvert_{_{x=L}} - <\overline{A_{v}v_{z}N^2_{z}}>\big\rvert_{_{x=0}}\right]\;dy 
\end{equation}

$\overline{J}_{NA_{1}}$ is associated with lateral (viscous) momentum dissipation, a contribution that becomes dominant in the presence of lateral boundary currents and strong vertical stratification. $\overline{J}_{NA_{2}}$ and $\overline{J}_{NA_{3}}$ are associated with surface wind stress and bottom frictional stress respectively, with non-vanishing contributions when these have non-zero (surface/bottom) curl, whereas the $\overline{J}_{NA_{4}}$ is associated with curvature in the vertical stratification.

As with the transport streamfunction, a trivial solution to (\ref{IntConstrain1}) for a stratified flow past finite amplitude topography is that when each integral term vanishes, implying an equal amount of PV being fluxed into and out of the control volume. In the expression for net lateral PV fluxes, bottom topography does not appear explicitly, and thus the vanishing of each integral does not require the integrand to behave in any particular way as a function of distance from the ridge crest. Nonetheless, the dissipative effects associated with lateral boundary currents will be localized to sloping bottom.

A resulting flow to consider is that in which the right hand side of (\ref{IntConstrain1}) is zero, meaning the net circulation can be described by lateral PV integral balances, just as in \citet{yang2000water}. This is the case of steady flow with vanishing \textit{Net Topographic PV Flux}, and is equivalent to restricting bottom topography to play the passive role of promoting horizontal divergence only, i.e. no frictional stresses act on the mean flow as it navigates the ridge.
%%%%%%%%%%%%%%%%%
%REFERENCES
%%%%%%%%%%%%%%%%%

\bibliographystyle{ametsoc2014}
\bibliography{PV_BBL.bib}

%%%%%%%%%%%%%%%%%
% TABLES
%%%%%%%%%%%%%%%%%

\begin{table}
\caption{Parameters relevant to lateral boundary currents. For stratified simulations, variables are vertically averaged within 100m from the bottom, localized to boundary currents, given the bottom intensified nature of the flows in all stratified simulations.}\label{table_one}
\begin{tabular}{l|ccccccccccc}
\hline\hline
Experiment &	$A_{_{H}}\;[m^2s^{-1}]$    &  $N^2_{0}[s^{-1}]$ & $\lambda_{M}\;[km]$	&	\multicolumn{2}{c}{$\overline{V}_{bot}[ms^{-1}$]} & \multicolumn{2}{c}{$|Ro|_{_{bot}} $}	 &\multicolumn{2}{c}{$Re$} &\multicolumn{2}{c}{$N_{bot}^2$} \\
\cline{5-12}
& & & &$_{\text{WBC}}$ & $_{\text{EBC}}$& $_{\text{WBC}}$ & $_{\text{EBC}}$ &$_{\text{WBC}}$ & $_{\text{EBC}}$&$_{\text{WBC}}$ & $_{\text{EBC}}$ \\
\hline\hline
\ \ $A_{fs}$&	$2500$&	 $--$&	$79.4$&	$0.45$&$0.345$&	$0.05$&$0.04$	&	$14.3$&$10.96$&--&--\\
\ \ $A_{ns}$&	$2500$&	$--$&	$79.4$&	$0.24$&$0.22$	&	$0.02$&$0.02$	&	$7.62$&$6.99$&--&--\\
\ \ $Ast_{fs}$& $2500$&   $10^{-5}$ &$79.4$ & $0.20$ & $0.28$ & $0.15$ & $0.09$ & $3.81$ & $8.26$ &3.06 &1.04 \\ 
\ \ $Ast_{ns}$& $2500$&   $10^{-5}$&$79.4$ & $0.11$ & $0.17$ & $0.08$ & $0.05$ & $2.86$ & $4.76$&3.07 & 0.87 \\\hline
\ \ $B_{fs}$ & $250$&	$--$ & $36.8$	&	$0.84$&$0.48$	&	$0.18$ & $0.1$	&	$123.6$&$70.7$&--&--\\
\ \ $B_{ns}$&	$250$&	$--$& $36.8$ &	$0.54$ &$0.43$	&	$0.12$ &$0.09$	&	$79.5$&$63.3$&--&--\\
\ \ $Bst_{fs}$& $250$&   $10^{-5}$ & $36.8$ & $0.47$ & $0.47$ & $0.45$ & $0.16$ & $41.2$ & $66.24$ & 2.39 & 0.85\\ 
\ \ $Bst_{ns}$& $250$&   $10^{-5}$& $36.8$ & $0.24$ & $0.36$ & $0.18$ & $0.12$ & $26.5$ & $53$ & 2.82 &  0.83\\ \hline
\ \ $C_{fs}$&	$25$& 	$--$&	 $17.1$ &	$0.94$&$0.48$	&	$0.44$    & $0.22$	&	$643$&$328.3$&--&--\\	
\ \ $C_{ns}$&	$25$& 	$--$&	 $17.1$&	$0.78$&$0.65$	&	$0.36$&$0.31$	&	$533.5$&$444.6$&--&--\\
\ \ $Cst_{fs}$& $25$&   $10^{-5}$ & $17.1$ & $0.47$ & $0.62$ & $0.72$ & $0.17$ & $218.9$ & $478.8$ &2.63 & 0.87\\ 
\ \ $Cst_{ns}$& $25$&   $10^{-5}$&$17.1$ & $0.32$ & $0.39$ & $0.28$ & $0.91$ & $164.2$ & $321.5$ & 2.2 & 0.7 \\ 
\hline\hline
\end{tabular}
\\\\
\footnotesize{$Ast_{fs}$, $Ast_{ns}$: Stratified simulation A performed using \textit{free slip} and \textit{no slip} tangential boundary condition. $A_{fs}$ refers to homogeneous (single density) simulation A using \textit{free slip} boundary condition.}\\
\footnotesize{$_{\text{WBC}}$, $_{\text{EBC}}$: Value calculated for the western/eastern boundary current, respectively.}
\end{table}

\begin{table}
\caption{Parameters associated with bottom boundary layer dynamics, arising on the region away from the influence from lateral boundary currents, along f/H contours atop the ridge, in all our simulations. The region is defined laterally as $\lambda_{M} < x < 250 - \lambda_{M}$. Values represent an average over this region.}\label{table_two}
\begin{tabular}{lcr}
\hline\hline
Parameter & Description &  Value	 \\
\hline\hline
$\theta$ & (small angle) topographic slope & $0.002$ \\ 
$A_{\nu}$ & Vertical mixing coefficient & $\approx5\times10^{-3} [m^2s]$ \\
$A_{\kappa}$ & Vertical diffusivity coefficient & $\approx5\times10^{-3} [m^2s]$ \\
$N_{c}^2$& Buoyancy frequency at ridge crest & $1.3\times10^{-5}\;[s^{-2}]$ \\
$S$ & Slope Burger number $S=(\theta N_{c}/f)^2$ & $0.003$ \\ 
$H_{p}$ & Prandtl depth & $55[m]$\\
$E$ & Ekman number  & $0.026$ \\
$P_{r}$ & Prandtl number & $\approx1$\\
$\mathcal{T}_{shut}$ & Buoyancy shutdown timescale & $8360$ [days] \\ 
$\mathcal{T}_{spin}$ & Spindown timescale & $0.57$ [days] \\
$\mathcal{T}_{diff}$ & Diffusive timescale & $ 3.5$ [days] \\
$\mathcal{T}_{adv}$ & Advective timescale & $ 3.8$ [days] \\
$\mathcal{T}_{inertial}$ & Inertial timescale & $0.58$\\
$\delta_{Ek}$ & Ekman layer thickness  & $9\;[m]$\\ 
$\delta_{T}$ & Diffusive boundary layer thickness & $55\;[m]$\\
$\omega_{bbl}$ & Eddy forcing frequency & $2.31\times10^{-6}\;[s^{-1}]$\\
\hline\hline
\end{tabular}
\end{table}

%%%%%%%%%%%%%%%%%
% Figures
%%%%%%%%%%%%%%%%%

\end{document}